\documentclass[12pt]{article}
\usepackage{jheppub}
\usepackage{lmodern}
\usepackage{a4wide,epsfig,psfrag,amsmath,amssymb,scalefnt}
\usepackage[dvipsnames]{xcolor}
\usepackage{amsmath,comment,braket}
\usepackage{placeins}
\usepackage{breqn}
\usepackage{slashed}
\usepackage{comment}
\usepackage{subcaption}
\usepackage{hyperref}
\usepackage{ifthen}
\usepackage{siunitx}
\usepackage{dsfont}
\usepackage{xfp}

\graphicspath{ {figs/},{diags/} }

\newcommand\db[1]{\ifthenelse{\equal{#1}{0}}{\Delta B=#1}{|\Delta B|=#1}}
\newcommand{\msbar}{\overline{\rm MS}}

\allowdisplaybreaks

\newcommand{\lt}{\left}
\newcommand{\rt}{\right}
\newcommand{\no}{\nonumber}
\newcommand{\nn}{\nonumber\\}

\newcommand{\eq}[1]{Eq.~(\ref{#1})}
\newcommand{\eqsand}[2]{Eqs.~(\ref{#1}) and (\ref{#2})}
\newcommand{\eqsto}[2]{Eqs.~(\ref{#1}) to (\ref{#2})}

\newcommand{\fig}[1]{Fig.~\ref{#1}}

\newcommand{\lqcd}{\Lambda_{\rm QCD}} 

\newcommand{\dg}{\ensuremath{\Delta \Gamma}}

\preprint{P3H-26-030, TTP26-014}

\title{
  Next-to-next-to-leading QCD corrections to the
  $\mathbf{B^+}$-$\mathbf{B_d^0}$,
  $\mathbf{D^+}$-$\mathbf{D^0}$,
  and   $\mathbf{D_s^+}$-$\mathbf{D^0}$
  lifetime ratios
}

\author[a]{Francesco Moretti,}
\author[a]{Ulrich Nierste,}
\author[a]{Pascal Reeck,}
\author[a]{Matthias Steinhauser}

\affiliation[a]{
  Institut f{\"u}r Theoretische Teilchenphysik,
  Karlsruhe Institute of Technology (KIT),
  Wolfgang-Gaede Stra\ss{}e 1, 
  76131 Karlsruhe, Germany
}

\abstract{
The total decay widths of heavy mesons can be systematically calculated
in terms of an expansion in the two parameters $1/m_Q$ and
$\alpha_s(m_Q)$, where $Q=c,b$ denotes the heavy quark. The dominant
contributions to meson lifetime splittings stem from terms which are
suppressed by $1/m_Q^3$ with respect to the leading universal
contribution to the total decay width. We calculate
three-loop contributions 
of order $\alpha_s^2/m_q^3$ to the lifetime ratios
$\tau(B^+)/\tau(B_d^0)$, $\tau(D^+)/\tau(D^0)$, and 
$\tau(D_s^+)/\tau(D^0)$ in the limit of exact isospin and V-spin
symmetry, respectively. Furthermore, we present new $\alpha_s/m_q^3$
corrections to the Cabibbo-suppressed terms in $\tau(B^+)/\tau(B_d^0)$. 
Combining our perturbative coefficients with hadronic matrix elements
calculated from Heavy Quark Effective Theory sum rules, we  find
$\tau(B^+)/\tau(B_d^0)= {1.072 \pm 0.024 }$. Using 
hadronic matrix elements from a recent lattice QCD calculation we find 
$\tau(D^+)/\tau(D^0)={2.344 \pm 0.170} $ and 
$\tau(D_s^+)/\tau(D^0)={1.289 \pm 0.042}$.
We find good agreement of our predictions with experimental data, which
constitutes a successful probe of the calculations of hadronic matrix
elements and permits estimates of the  unknown $1/m_Q^4$
contributions as well as the V-spin breaking terms in
$\tau(D_s^+)/\tau(D^0)$.
}
  
\begin{document}

\maketitle
\flushbottom





\section{Introduction}
Lifetimes of heavy hadrons $H_Q$ can be calculated with the \emph{Heavy Quark
  Expansion (HQE)}, which is an operator product expansion with the
heavy quark mass $m_Q$ as the hard scale \cite{Khoze_1983,Shifman:1984wx,Khoze:1986fa,Shifman:1986mx,Bigi:1992su}. The HQE results in expressions
of the schematic form 
\begin{align}
 \Gamma(H_Q) &= 
  \Gamma_3 
  { \langle {\cal O}_3\rangle}
  +  
  \Gamma_5 \frac{\langle {\cal O}_5\rangle}{m_Q^2} 
  +
{
 \Gamma_6 \frac{\langle {\cal O}_6\rangle}{m_Q^3} 
  +
 \Gamma_7 \frac{\langle {\cal O}_7\rangle}{m_Q^4} 
               + \ldots }
\nn
               &\qquad \qquad \qquad +
16 \pi^2 \lt[ 
\tilde{\Gamma}_6 \frac{\langle \tilde{\cal O}_6\rangle}{m_Q^3} 
+
\tilde{\Gamma}_7 \frac{\langle \tilde{\cal O}_7\rangle}{m_Q^4} 
+ \ldots \rt]
                \label{eq:hqe}
\end{align}
for the total width $\Gamma(H_Q)$.  Here
$\langle {\cal O}_n\rangle\equiv \bra{H_Q} {\cal O}_n \ket{H_Q}$ denotes
a hadronic matrix element of a local operator of dimension $n$ and
$\Gamma_n$, $\tilde\Gamma_n$ represent Wilson coefficients which are
calculable in perturbation theory to the desired order in the coupling
constant $\alpha_s$ of Quantum Chromodynamics (QCD). The first term
$\Gamma_3 \braket {{\cal O}_3}$ equals the QCD-corrected quark decay
rate and is thus a universal contribution, entailing that to first
approximation all $b$-flavoured hadrons have the same lifetime.  Higher
dimensions $n$ corresponds to higher powers of the QCD scale parameter
$\lqcd$, so that the HQE expansion parameters are $\lqcd/m_b\sim 0.1$
and $\lqcd/m_c\sim 0.3$ for the two heavy-quark systems of interest.
The Wilson coefficients involve $\alpha_s(m_b)\sim 0.2$ and
$\alpha_s(m_c)\sim 0.3$, respectively. Whether the expansions in these
two parameters actually converge to a satisfactory level can only be
found out if $\Gamma(H_Q)$ is calculated to sufficiently high orders in
these parameters.

An important feature of \eq{eq:hqe} is the appearance of the numerical enhancement
factor $16 \pi^2$ in the second line of \eq{eq:hqe}, emerging from
contributions in which a valence  quark in $H_Q$ participates in the
weak decay amplitude. While $\Gamma_3$ stems from
three-body decays like $b\to c \bar u d$, $\tilde{\Gamma}_6(B_d^0)$  
receives  contributions from e.g.\ $\bar b d \to c \bar u$ and the
two-particle phase space comes with one power of $1/(16 \pi^2) $ less
than the terms in the first line of \eq{eq:hqe}.
There are no terms with enhancement factors of higher powers of
$16 \pi^2$, so that the terms in the second line of \eq{eq:hqe} do not spoil the convergence of the HQE.
However, in the case of charmed hadrons one has $16 \pi^2 \lqcd^3/m_c^3
\sim 1$, so that lifetime differences of order 1 are allowed.

Lifetime differences emerge from the feature that the matrix elements of
operators with $n\geq 5$ are different for different hadrons $H_q$.  The
lifetime splittings in the isospin doublets $(B^+,B_d^0)$ and
$(\Xi_b^0,\Xi_b^-)$ and their counterparts in the charm system are
especially interesting, because isospin symmetry simplifies the
calculation in two ways:
\begin{itemize}
\item[(i)] The hadronic matrix elements  of isospin
  partners are related to each other in a trivial way. In particular, 
  the terms in the first line of   \eq{eq:hqe} drop out from the
  lifetime splittings in doublets in the symmetry limit. Since strong isospin is an excellent
  symmetry of QCD, holding with an accuracy of  2\%, isospin-breaking
  effects in the hadronic matrix elements can be safely neglected and the lifetime differences solely
  stem from the terms in the second line of  \eq{eq:hqe}.  
\item[(ii)] The operators $\tilde{\cal O}_6$ are four-quark operators 
  which distinguish between e.g.\ $B^+$ and $B_d^0$, for example
  $\bra{B_d^0} \bar b \gamma_\mu d \, \bar d \gamma^\mu b \ket{B_d^0}
  \neq \bra{B^+} \bar b \gamma_\mu d \, \bar d \gamma^\mu b
  \ket{B^+}$, so that they give a non-zero contribution to the $B^+$-$B_d^0$
  lifetime difference. The final expression for this quantity will only
  contain isospin-violating operators and these operators cannot mix
  into the isospin-conserving operators  ${\cal O}_n$. In particular,
  one needs no counterterms proportional to the lower-dimensional operators
  ${\cal O}_{3}$ or ${\cal O}_{5}$, which are currently show-stoppers to a reliable calculation of
  the dimension-6 operator matrix elements $\braket{\tilde{\cal O}_6}$ with lattice QCD.
\end{itemize}
These points also apply to V-spin symmetry, which corresponds to unitary
rotations of the $(u,s)^T$ quark field doublet, with the caveat that
V-spin symmetry is broken at a $\sim 30\%$ level due to $m_s - m_u \sim
0.3\,\lqcd$. Thus we expect a larger deviation of our prediction of the
$D_s^+$-$D^0$  lifetime ratio from data than in the cases of isospin
doublets.

Theoretical predictions of lifetime ratios are related to calculations
  of the differences $\Gamma(H_Q)-\Gamma(H_Q^\prime)$ of the total
  decay rates of the considered hadrons via
\begin{align}  
  \frac{\tau(H_Q^\prime)}{\tau(H_Q)} -1
  &=\, \tau^{\mathrm{exp}}(H_Q^{\prime}) \lt(  \Gamma(H_Q)-\Gamma(H_Q^\prime)  \rt).
\label{eq:ltr}
\end{align}  
Here ``exp'' means that the experimental value of the lifetime is used
in the prefactor. $\Gamma(H_Q)$ is related to the forward matrix element
$H_Q \to H_Q$ through the optical theorem. 
Specifying the discussion to $\tau(B^+)/\tau(B_d^0)$ now, the
  Cabibbo-favoured contribution to this lifetime ratio is proportional
  to $|V_{ud} V_{cb}|^2$.
For the computation of $\Gamma(H_b)$ one first integrates out all heavy degrees of freedom with masses of the order of the electroweak scale. This results in the  $\db1$ Hamiltonian $\mathcal{H}^{\db1}_{\rm eff}$ which describes the weak decays of $b$-flavoured hadrons in the SM, where $B$ is the beauty quantum number. The low-energy dynamics, related to energy scales of order $m_b$ and below, is described by effective dimension-6 operators. These $\db1$ operators are multiplied by  Wilson coefficients describing the 
short-distance dynamics, which can be calculated in perturbation theory. The $\db1$
Wilson coefficients are known up to next-to-next-to-leading order (NNLO) of QCD \cite{Gorbahn:2004my}.
The implementation of the HQE for the calculation of $\tilde \Gamma_n$ amounts to the matching of the $\bar b q\to \bar b q$ forward matrix element calculated with $\mathcal{H}^{\db1}_{\rm eff}$ to  effective local $\db0$ operators, where $q$ is a valence quark of $H_b$. The perturbative calculation of the $\db0$  Wilson coefficients is justified by $m_b \gg \lqcd$.

In processes like $\bar b \to \bar c u \bar d$ with four different quark flavours 
$\mathcal{H}^{\db1}_{\rm eff}$ comprises only 
two current-current operators, which describe the $W$-mediated SM tree diagram including QCD corrections. 
For decays into a $u\bar u$ or $c\bar c$ pair one also encounters penguin operators, with much smaller Wilson coefficients. Furthermore, in our calculation of $\tau(B^+)/\tau(B_d^0)$ all penguin effects are CKM-suppressed 
by two powers of the Wolfenstein parameter $\lambda\sim 0.22$, so that no NNLO precision is currently needed for the latter. 
The CKM-leading contribution to  $\tau(B^+)/\tau(B_d^0)$ has been calculated to NLO in Ref.~\cite{Beneke:2002rj,Franco:2002fc}. Judging the accuracy of the result by the dependences on renormalisation scale 
and scheme,  one concludes that an NNLO calculation is needed to confront today's experimental result  \cite{HFLAV:2024ctg} with theory. This calculation is especially timely, because meanwhile hadronic matrix elements have been calculated with QCD sum rules \cite{Kirk:2017juj,Black:2024bus} and also first complete lattice-QCD computations of all matrix elements entering $\tau(D^+)/\tau(D^0)$ and $\tau(D_s^+)/\tau(D^0)$ are available~\cite{Black:2026dzp,Black:2026rbz}.\footnote{First exploratory lattice calculations were presented in Refs.~\cite{DiPierro:1998ty,DiPierro:1999tb,Becirevic:2001fy}; for a first determination of bare matrix elements with domain wall fermions see Ref.~\cite{Lin:2022fun}.}

\begin{figure}[t]
    \centering
    \begin{tabular}{cc}
         \includegraphics[width=0.45\linewidth]{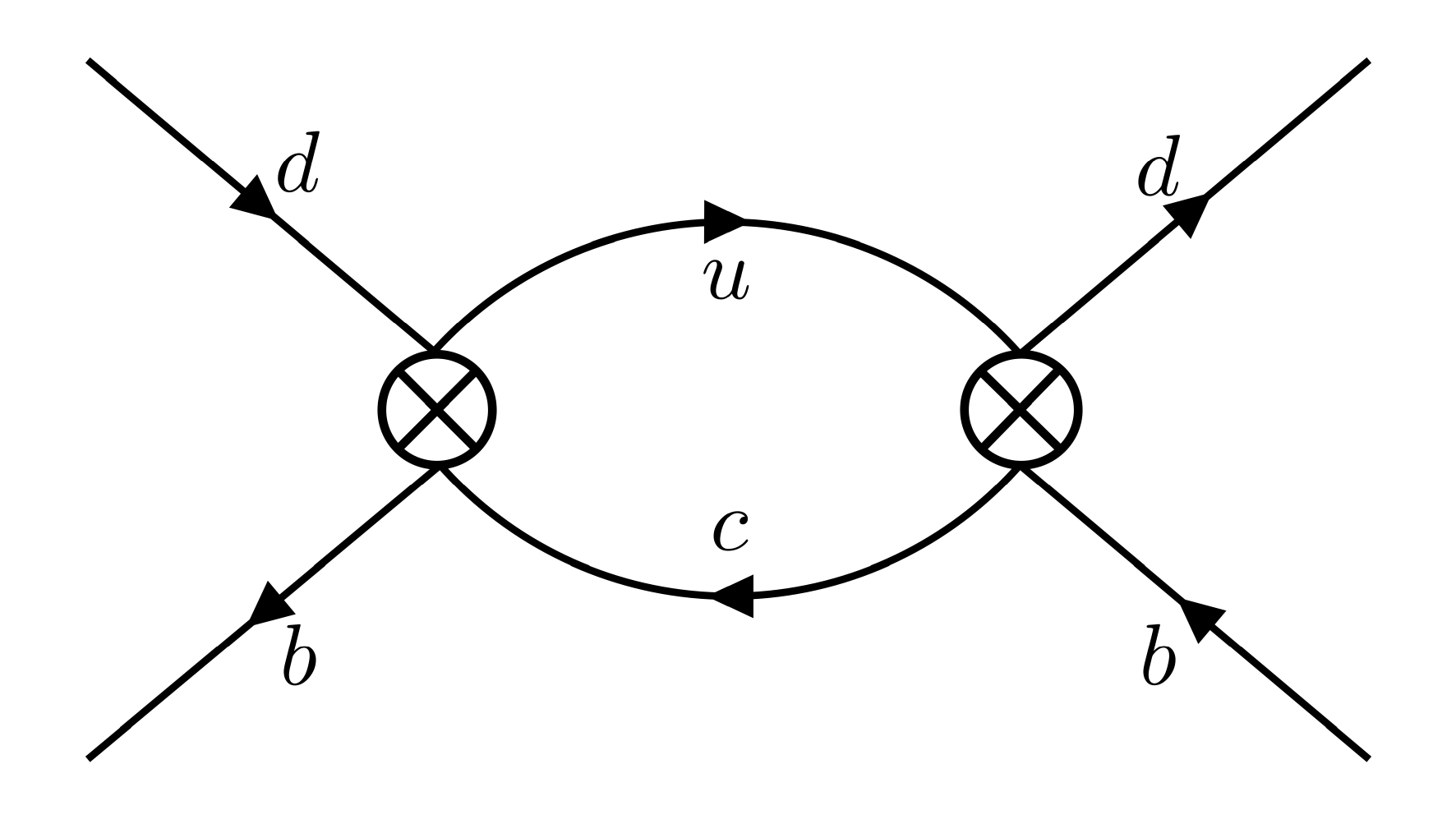}
         &\includegraphics[width=0.45\linewidth]{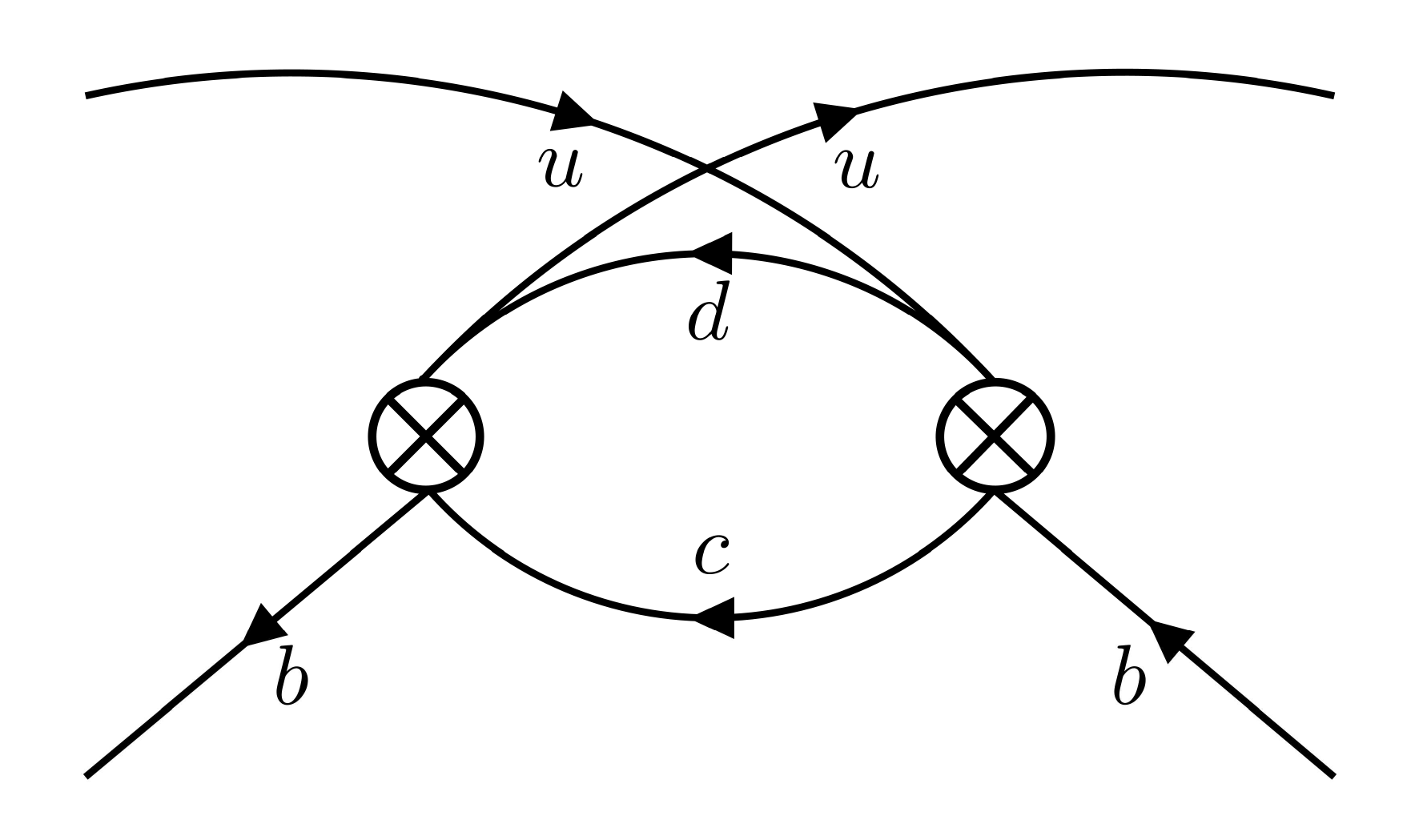}
    \end{tabular}
    \caption{Weak annihilation (left) and Pauli interference (right) diagrams in the leading order of QCD. They contribute to $\Gamma(B_d^0)$ and $\Gamma(B^+)$, respectively. The crosses represent the insertion of $|\Delta B| = 1$ operators from the effective Hamiltonian \eqref{eq:eff-db1-ham}. CKM-suppressed contributions are not shown. }
    \label{fig:wa-pi-LO}
\end{figure}

The dominant contribution to  $\tau(B^+)/\tau(B_d^0)-1$ stems from  {\it Weak  annihilation (WA)} and the {\it Pauli interference (PI)} diagrams illustrated in Fig.~\ref{fig:wa-pi-LO} at LO. They contribute to 
neutral and charged $B$ meson decay rates, respectively. The new results of this paper are
\begin{itemize}
\item[(i)] NNLO corrections to the WA and PI contributions, i.e.\ the CKM-favoured piece of $\tau(B^+)/\tau(B_d^0)-1$,
\item[(ii)] NLO corrections to the CKM-suppressed WA and PI contributions with two current-current operators, and
\item[(iii)]  the application of the results to the $D^+$-$D^0$ and $D_s^+$-$D^0$ lifetime differences. 
\end{itemize}
Conceptually, the calculation is similar to the one of the decay matrix element $\Gamma_{12}^q$ of 
$B_q^0$-$\bar B_q^0$ mixing, which determines the width differences $\dg_q$ among the mass eigenstates 
and the CP asymmetries in flavour-specific decays. Just as in the case of $\tau(B^+)/\tau(B_d^0)$ 
the starting point was an NLO result for the current-current contribution \cite{Beneke:1996gn, Beneke:2003az,Lenz:2006hd}
and a systematic calculation of sub-leading contributions \cite{Gerlach:2022wgb,Gerlach:2025tcx} up to the complete NNLO prediction\cite{Nierste:2025muk} lead to a substantial reduction of the scale and
scheme dependences of the predicted  $\Gamma_{12}^q$. Another NNLO calculation in this context has addressed the 
term $\Gamma_3$ in \eq{eq:hqe}, which constitutes the universal piece common to all decay rates of 
$b$-flavoured hadrons \cite{Egner:2024azu,Egner:2024lay}. 

\section{Effective theories}
\label{sec::eft}
\begin{figure}[t]
    \centering
    \begin{tabular}{cc}
         \includegraphics[width=0.45\linewidth]{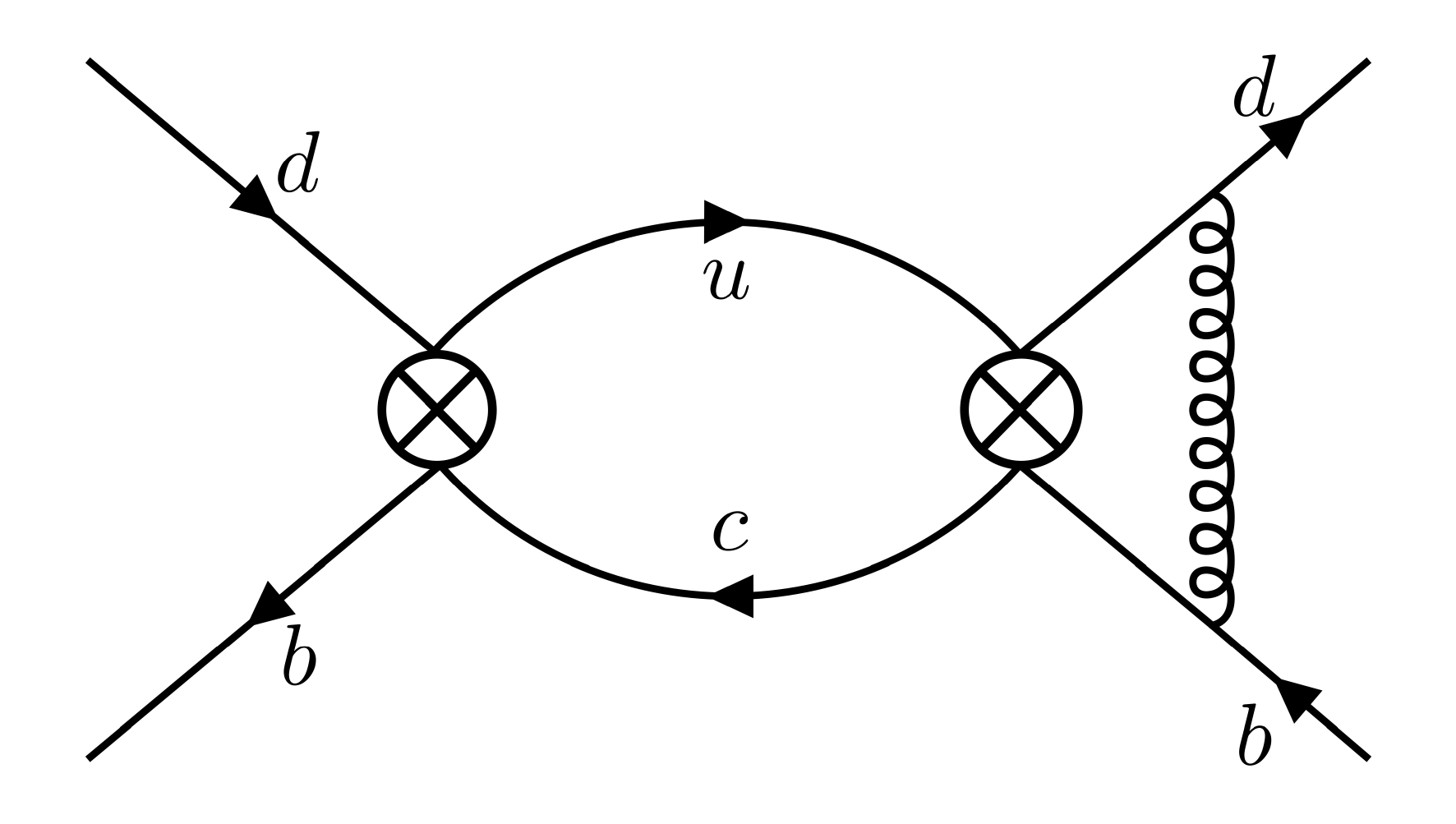}
         &\includegraphics[width=0.45\linewidth]{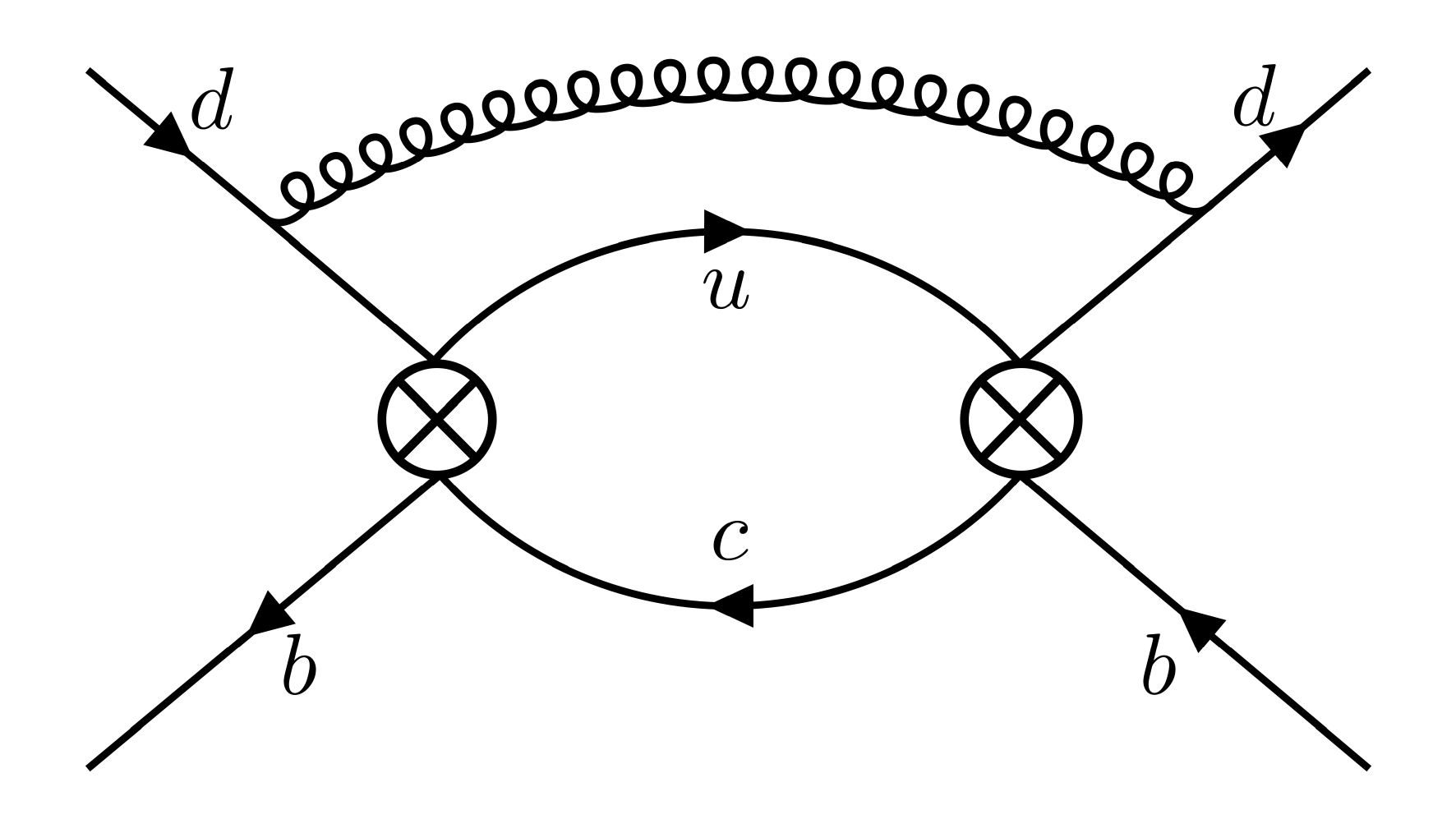}\\
         \includegraphics[width=0.45\linewidth]{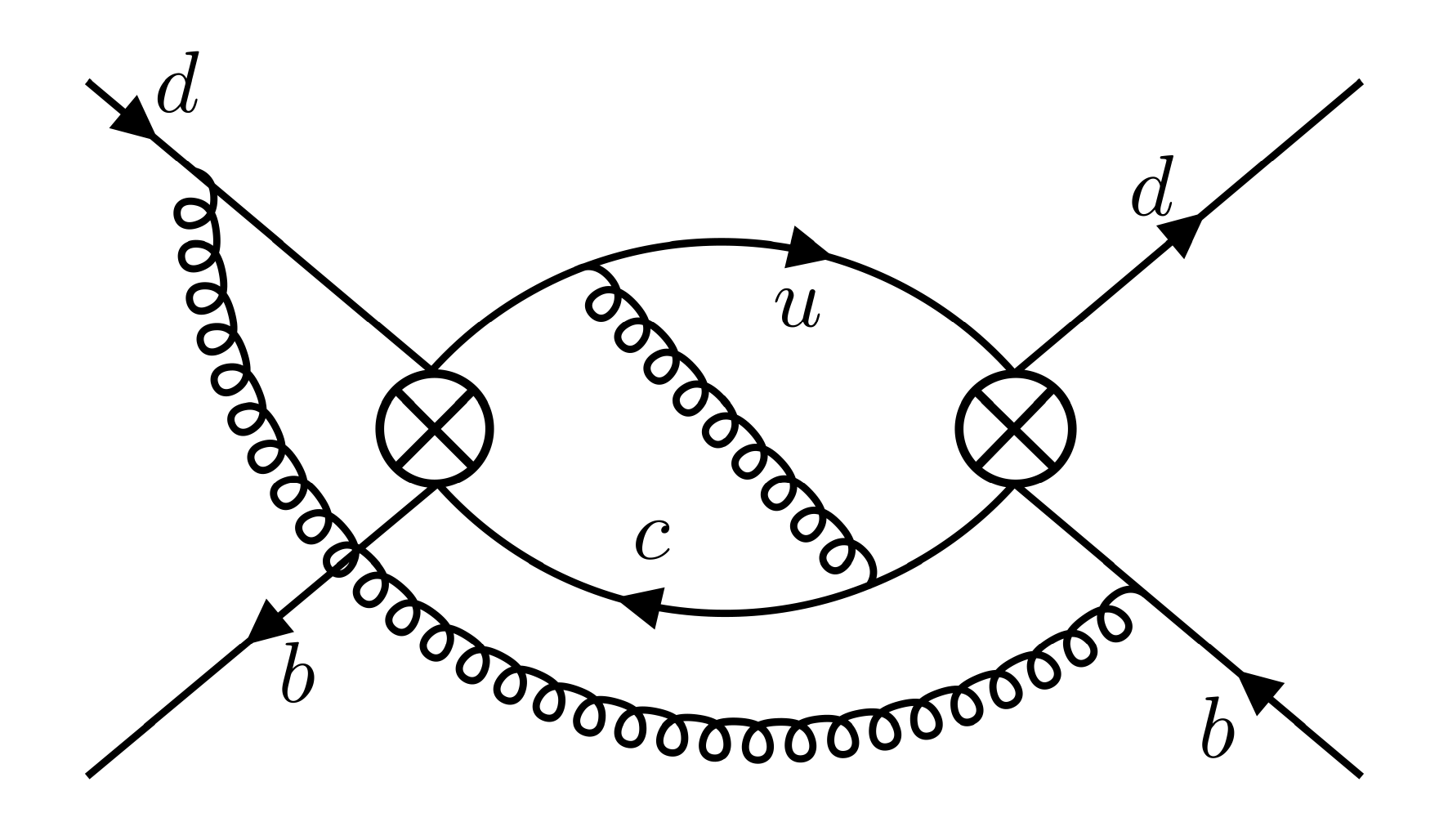}
         &\includegraphics[width=0.45\linewidth]{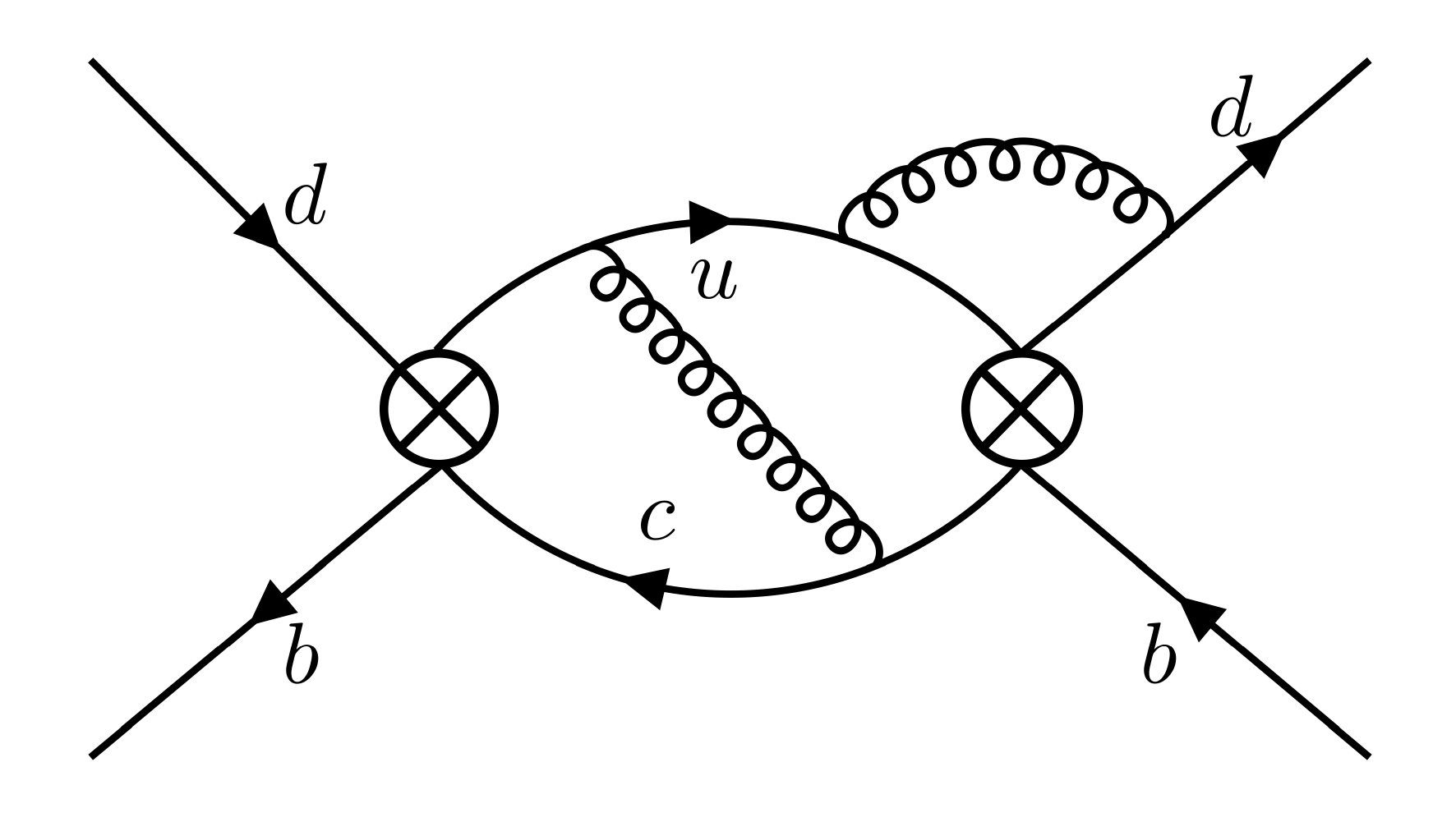}
    \end{tabular}
    \caption{Sample Feynman diagrams contribution to $\Gamma(B_d^0)$
    at NLO (top) and NNLO (bottom). Analogue diagrams also exist for the PI contribution.}
     \label{fig:wa-NLO-NNLO}
\end{figure}

The optical theorem expresses the total decay width of $H_b$ as
\begin{eqnarray}
\Gamma(H_b) &=& \frac{1}{2 M_{H_b}}
  \langle H_b| \mathcal{T}
  |H_b\rangle \nn 
\mbox{with}\qquad\quad  
\mathcal{T}&=&\mbox{Im}\; i \int\textrm{d}^4 x\;
     T\left[\mathcal{H}^{\db1}_{\rm eff}(x)\mathcal{H}^{\db1}_{\rm eff}(0)\right]  
\label{eq:decay-width}     
\end{eqnarray}
and ``Im'' in the definition of transition operator
understood to be applied to the matrix element $\langle H_b| \ldots |H_b \rangle $. 

Sample Feynman diagrams 
for the decay of a neutral and charge $B$ meson
to LO are given in Fig.~\ref{fig:wa-pi-LO}. For illustration we show
in Fig.~\ref{fig:wa-NLO-NNLO} Feynman diagrams which contribute to $\Gamma(B_d^0)$ at NLO and NNLO.
The following step is to perform a Heavy Quark Expansion (HQE) of the transition operator in powers of $m_b\gg \Lambda_{\rm QCD}$. This amounts to expressing the bi-local correlator in Eqs.~\eqref{eq:decay-width} and~\eqref{eq:decay-width2} in terms of {\it local} $\db0$ operators multiplied by the corresponding matching coefficients. While the matrix elements of such operators need to be evaluated using non-perturbative methods, e.g. lattice QCD or sum rules, the matching coefficients are determined in perturbation theory, and depend on the $\db1$ Wilson coefficients, the strong coupling constant $\alpha_s$ and the quark masses $m_b$ and $m_c$. Higher dimensional operators correspond to increasingly suppressed terms in the HQE.
The main scope of this work is the evaluation of the matching coefficients to NNLO accuracy in QCD. Effectively, this corresponds to a three-loop calculation on the $\db1$ side, and a two-loop calculation on the $\db{0}$ side. The methodology of the matching calculation is described in Refs.~\cite{Neubert:1996we,Beneke:1996gn,Beneke:1998sy,Ciuchini:2001vx,Franco:2002fc,Beneke:2002rj,Beneke:2003az}.

$\mathcal{H}^{\db1}_{\rm eff}$ will be introduced in Sec.~\ref{subs:db1}.  
The HQE of \eq{eq:hqe} expresses the bilocal matrix element $\langle H_b| \mathcal{T}
  |H_b\rangle$ in terms of local $\Delta B=0$ operators, to be discussed in Sec.~\ref{subs:db0}.

\subsection[${|\Delta B|=1}$ hamiltonian]{$\mathbf{|\Delta B|=1}$ hamiltonian\label{subs:db1}}
For convenience, we adopt the  traditional operator basis~\cite{Gilman:1979bc,Buras:1989xd} with the $\db1$ Hamiltonian
\begin{align}
    \mathcal{H}^{\db1}_{\rm eff} = 4\frac{G_F}{\sqrt{2}}
    \sum_{q_{3}=d,s}\Bigg\{\sum_{q_{1,2}=u,c}
    \lambda_{q_1q_2q_3}&\left[C_1(\mu_1)Q_1^{q_1q_2q_3}(\mu_1) + C_2(\mu_1)Q_2^{q_1q_2q_3}(\mu_1) \right] \nonumber\\
    &-\lambda_{t}^{q_3}\sum_{i=3}^6 C_i(\mu_1)Q^{q_3}_i(\mu_1)\Bigg\}+ h.c.,\label{eq:eff-db1-ham}
\end{align}
where $\lambda_{q_1q_2q_3}=V_{q_1 b}^* V_{q_2q_3}$ and $
\lambda_t^q=V_{tb}^* V_{tq_3}$ comprise the CKM matrix elements 
and $C_i$ $(i=1,\ldots,6)$ are the Wilson coefficients encoding 
the short-distance physics associated with energies above the 
renormalisation scale $\mu_1$. The current-current operators
\begin{eqnarray}
    & Q_1^{q_1q_2q_3} = (\overline{b}^i \gamma^\mu P_L q^j_1)\;( \overline{q}_2^j \gamma_\mu P_L q_3^i)\label{eq:q1}\,,\nonumber\\
    & Q_2^{q_1q_2q_3} = (\overline{b}^i \gamma^\mu P_L q_1^i)\; (\overline{q}_2^j \gamma_\mu P_L q_3^j)\label{eq:q2}\,,
\end{eqnarray}
where the summation over the colour indices $i$ and $j$ is understood, are accompanied with Wilson coefficients $C_{1}$
and $C_2$ of order 1, while those of the four-quark penguin operators in the second line of \eq{eq:q2} are much smaller. In our NNLO calculation we restrict ourselves to the CKM-favoured 
contributions, which strictly only involve the CKM factor $|\lambda_{cud}|^2$. 
In the PI diagrams $|\lambda_{cud}|^2$ trivially combines with the CKM-suppressed contribution proportional to  
$|\lambda_{cus}|^2$ to an expression found by replacing $V_{ud}\to 1$ in the CKM-favoured piece, so that we
include this piece as well. (The WA contribution  does not involve 
$\lambda_{cus}$.)

\begin{figure}[tp]
    \centering
    \begin{minipage}[t]{0.45\textwidth}
        \centering
        \includegraphics[scale=0.25]{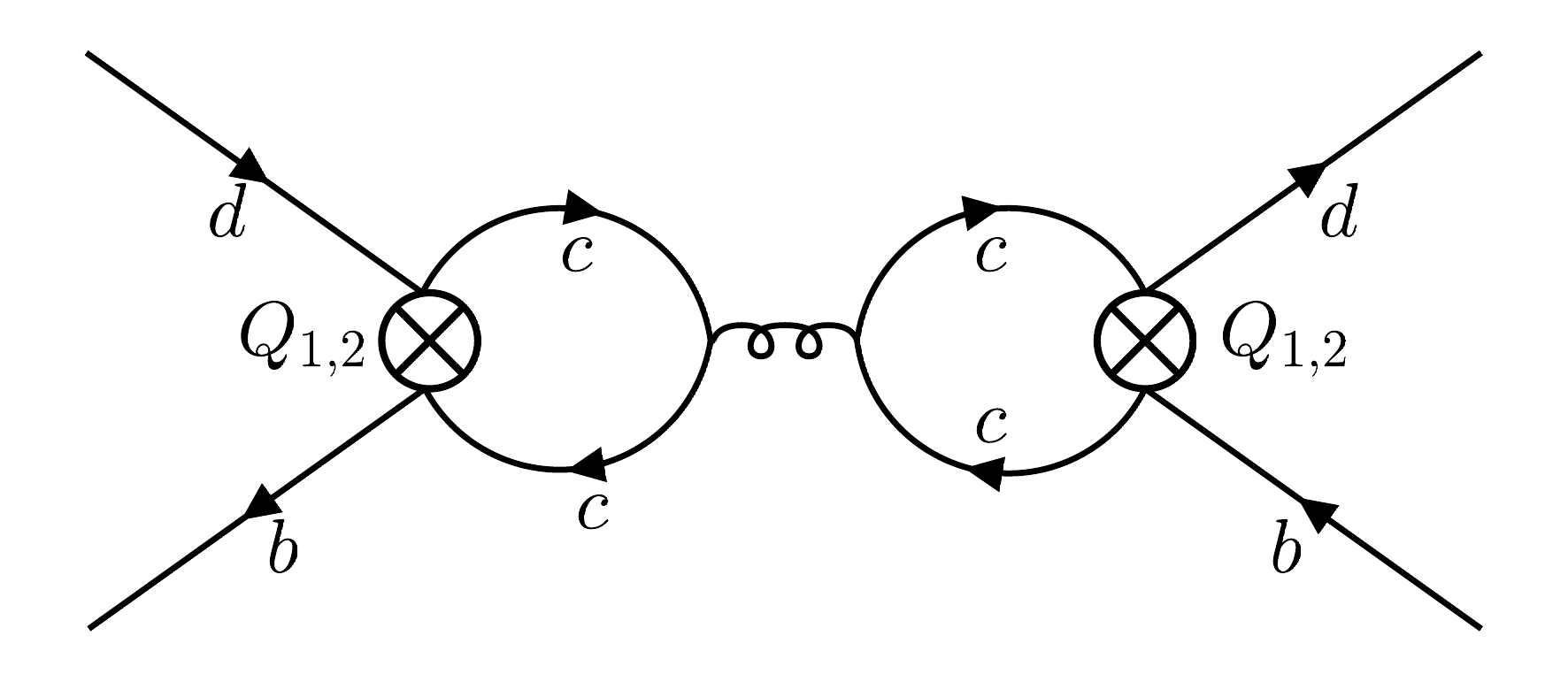}
         \caption{Penguin-like contribution at NLO to the decay width involving an intermediate $c\bar c$ pair in the hard loops. These contribution are CKM-suppressed by a factor $|V_{cd}|^2/|V_{ud}|^2$ with respect to the CKM-favoured weak annihilation process. Diagrams with internal $u$ quark are not shown
         but included in our analysis.}
         \label{fig:NLO-Peng-Like}
    \end{minipage}%
    \hspace{1cm}
    \begin{minipage}[t]{.45\linewidth}
        \centering
        \includegraphics[scale=0.19]{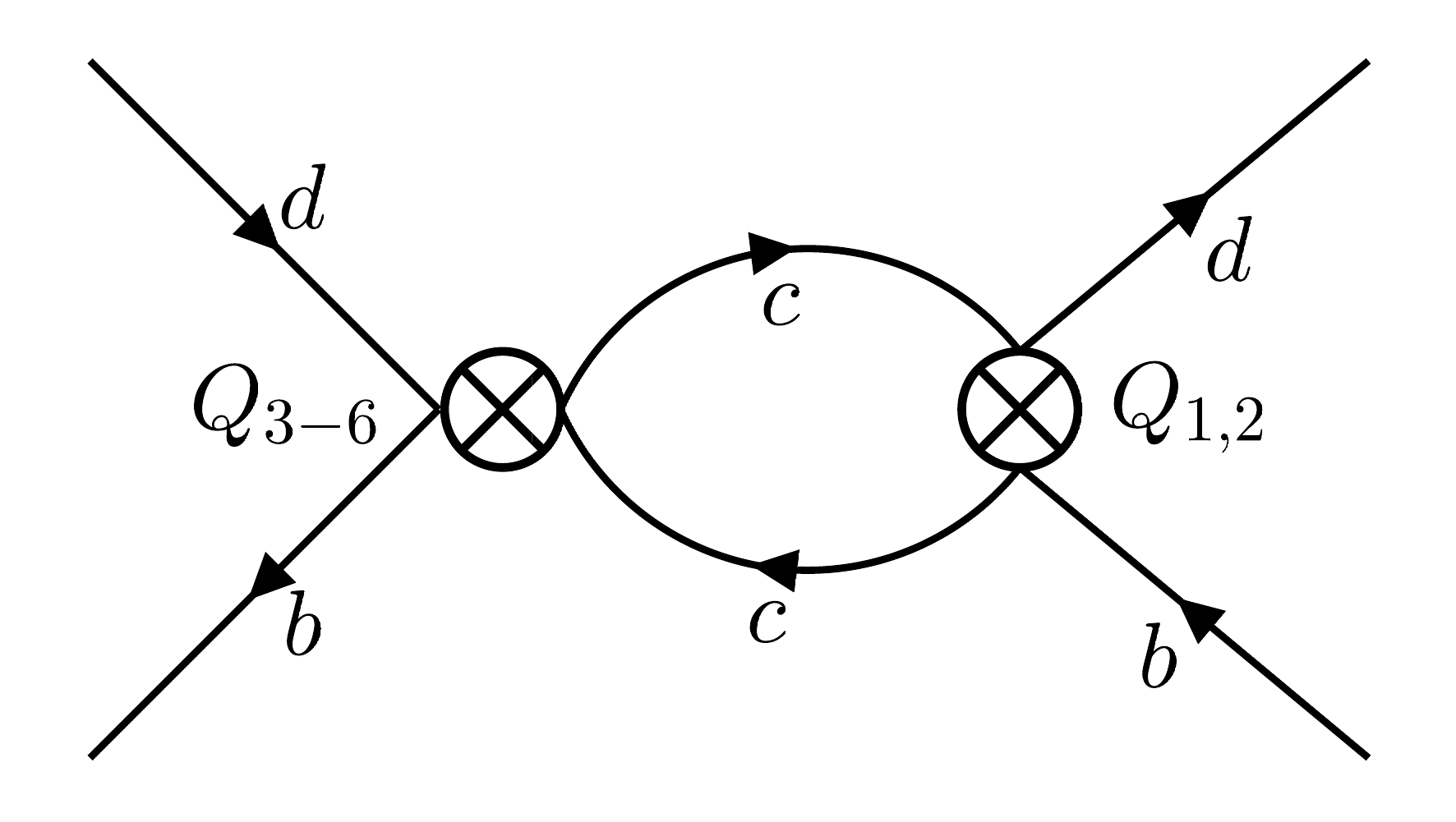}
        \caption{Weak annihilation diagram involving one penguin operator $Q_{3-6}$. These are needed at NLO as counterterms to the CKM-suppressed part of the decay width. Diagrams with internal $uu$ pairs are not shown
        but included in our analysis.}
        \label{fig:LO-Peng}
    \end{minipage}
\end{figure}

In addition, we extend the work in Ref.~\cite{Beneke:2002rj} and compute the CKM-suppressed contributions proportional to $|\lambda_{ccd}|^2$,
$\lambda^\prime_{cud}$ and $|\lambda_{uud}|^2$ to NLO accuracy. 
This includes diagrams as in Fig.~\ref{fig:wa-pi-LO} with internal $uc$ quarks for the WA case and with internal $ud$ quarks in the PI case, both at one- and two-loop order.
Furthermore, we have to consider one-particle reducible diagrams as in Fig.~\ref{fig:NLO-Peng-Like}.
They lead to CKM factors $|\lambda_{ccd}|^2$ and $|\lambda_{uud}|^2$ but also to
\begin{align}
\lambda^\prime_{cud}\equiv2{\rm Re}\left[V_{ub}^*V_{ud}V_{cb}V_{cd}^*\right]
=&\, -2\, |\lambda_{ccd}|^2 \bar{\rho} \, \lt[ 1 +{\cal O} (\lambda^4)\rt] \,,
\end{align}
which first appear at NLO accuracy. $\lambda^\prime_{cud}$ comes from diagrams of the form of Fig.~\ref{fig:NLO-Peng-Like}, where both a $cc$ and a $uu$ pair appear in the two loops. For the renormalisation of penguin-like contributions  we need counterterms proportional to
the penguin operators
\begin{align}
    \label{eq:penguin-historical}
    &Q^{q_3}_3=(\overline{b}^i \gamma^\mu P_L q_3^i)\;\sum_q ( \overline{q}^j \gamma_\mu P_L q^j),\nonumber\\
    &Q^{q_3}_4=(\overline{b}^i \gamma^\mu P_L q_3^j)\;\sum_q ( \overline{q}^j \gamma_\mu P_L q^i),\nonumber\\
    &Q^{q_3}_5=(\overline{b}^i \gamma^\mu P_L q_3^i)\;\sum_q ( \overline{q}^j \gamma_\mu P_R q^j),\nonumber\\
    &Q^{q_3}_6=(\overline{b}^i \gamma^\mu P_L q_3^j)\;\sum_q ( \overline{q}^j \gamma_\mu P_R q^i).
\end{align}
They enter LO diagrams
as shown in  Fig.~\ref{fig:LO-Peng}. Furthermore, 
we include the small Wilson coefficients $C_{3-6}$ at LO through the diagram of Fig.~\ref{fig:LO-Peng} as needed to cancel the renormalisation scale dependence. The described penguin effects have been calculated in the context of the Cabibbo-favoured piece of the lifetime ratio $\tau(B_s^0)/\tau(B_d^0)$ in Ref.~\cite{Keum:1998fd}, where only the  contribution with two charm loops in Fig.~\ref{fig:NLO-Peng-Like} was needed.

\begin{figure}[t]
    \centering
    \includegraphics[width=1\linewidth]{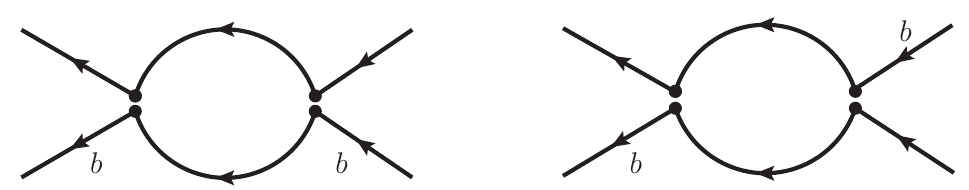}\\[-1.8cm]
    \hspace{0.4cm}
    \begin{Huge}\boldmath$\Longrightarrow$\end{Huge}
    \vspace{1.8cm}
    \caption{
    Diagrammatic description of the effect of inserting a Fierz-transformed operator in one of the two vertices, here the vertex on the right. The diagram on the l.h.s.~contains a spinor line where $b$-spinor are absent, such that EOMs cannot be used to simplify the expressions in the calculation. On the other hand, the r.h.s.~diagram will generate structures with one $b$-spinor on each fermion line. }
    \label{fig:fierz-explicit-PI}
\end{figure}

The main motivation to work
in the traditional basis is that we want to have one $b$ spinor on each external fermion line such that we can apply the equations of motion (EOMs) $\slashed{p}_b b = m_b b$,
where $p^\mu_b$ is the external $b$ quark four-momentum,
to simplify all the tensor structures. 
In the original version (cf. Fig.~\ref{fig:wa-pi-LO}) this is not the case, but we can achieve 
the desired expression if we replace one $|\Delta B|=1$ operator in the diagram with its Fierz-transformed 
version. For e.g.\ $Q_1^{q_1q_2q_3}$ this amounts to 
\begin{equation}
    \label{eq:fierz-transformation}
    Q_1^{q_1q_2q_3}=(\overline{b}^i \gamma^\mu P_L q^j_1)\;( \overline{q}_2^j \gamma_\mu P_L q_3^i)
    \stackrel{F}{=}
    (\overline{b}^i \gamma^\mu P_L q^i_3)\;( \overline{q}_2^j \gamma_\mu P_L q_1^j)=Q^{q_3q_2q_1}_2.
\end{equation}
Since the Fierz transformation only holds for $d=4$, a priori one expects different 
Wilson coefficients of $Q_j^{q_1q_2q_3}$ and its Fierz transform beyond LO. In Refs.~\cite{Buras:1989xd,Herrlich:1994kh,Asatrian:2017qaz,Egner:2024azu} it has been worked out how this is remedied 
through an appropriate definition of evanescent operators. As an advantage of the traditional 
operator basis, the Fierz transform simply interchanges  $Q_1^{q_1q_2q_3}$ and $Q_2^{q_3q_2q_1}$, 
which simplifies the calculation. 
We apply the Fierz transformation to the right operator in 
our diagrams
and obtain for the decay width
\begin{align}
    \label{eq:fierzed-decay-width}
    \widetilde{\Gamma}^{q_1q_2q_3} = \frac1{2M_{H_b}}&\sum_{i,j=1,2} \left(\frac{G_F}{\sqrt{2}}\right)^2\;
    |\lambda_{q_1q_2q_3}|^2
    \nonumber\\
    &\widetilde C_i(\mu_1) C_j (\mu_1) \; \langle H_b|\mbox{Im}\;i\int\mbox{d}^4x \;e^{i qx}\;T\left[Q_i^{\dagger
    q_3q_2q_1
    }(x)Q_j^{q_1q_2q_3}(0)\right]|H_b\rangle,
\end{align}

where $\widetilde C_i(\mu_1)$ is the Wilson coefficient of the Fierz-transformed operator $Q_i^{q_3 q_2 q_1}$. The forward scattering matrix defined by $\widetilde\Gamma^{q_3q_2q_1}$ leads to amplitudes where a $b$-spinor is present in each fermion line as illustrated in Fig.~\ref{fig:fierz-explicit-PI}, where the effect of the Fierz transformation is illustrated.

For the Wilson coefficients 
of the current-current operators we have the simple relation
\begin{eqnarray}
\label{eq:fierz-relation-WC}
    &C_1 = \widetilde C_2\nonumber\\
    &C_2 = \widetilde C_1,
\end{eqnarray}
with  a judicious choice for the evanescent operators
\begin{align}
    &E^{(1),q_1q_2q_3}_1=(\bar{b}^i\gamma^{\mu_1\mu_2\mu_3}P_L q_1^j)(\bar q_2^j \gamma_{\mu_1\mu_2\mu_3}P_L q_3^i)-(16-4\epsilon +{\cal A}_2 {\epsilon^2})Q_1^{q_1 q_2 q_3}\nonumber,\\
    &E^{(1),q_1q_2q_3}_2=(\bar{b}^i\gamma^{\mu_1\mu_2\mu_3}P_L q_1^i)(\bar q_2^j \gamma_{\mu_1\mu_2\mu_3}P_L q_3^j)-(16-4\epsilon +{\cal A}_2{\epsilon^2})Q_2^{q_1 q_2 q_3}\nonumber,\\
    &E^{(1),q_1q_2q_3}_1=(\bar{b}^i\gamma^{\mu_1\mu_2\mu_3\mu_4\mu_5}P_L q_1^j)(\bar q_2^j \gamma_{\mu_1\mu_2\mu_3\mu_4\mu_5}P_L q_3^i)-(256-224\epsilon +{\cal B}_1 {\epsilon^2})Q_1^{q_1 q_2 q_3}\nonumber,\\
    &E^{(1),q_1q_2q_3}_2=(\bar{b}^i\gamma^{\mu_1\mu_2\mu_3\mu_4\mu_5}P_L q_1^i)(\bar q_2^j \gamma_{\mu_1\mu_2\mu_3\mu_4\mu_5}P_L q_3^j)-(256-224\epsilon +{\cal B}_2{\epsilon^2})Q_2^{q_1 q_2 q_3},
    \label{eq:ev-db1}
\end{align}
where $\gamma^{\mu_1\cdots\mu_n}$ is a shorthand notation for $\gamma^{\mu_1}\cdots \gamma^{\mu_n}$. 
The coefficients of the $\epsilon^2$ terms first enter the calculation at NNLO and read~\cite{Egner:2024azu}
\begin{equation}
    {\cal A}_2=-4,\qquad {\cal B}_1=-\frac{45936}{115},\qquad {\cal B}_2=\frac{115056}{115}.
\end{equation}
For all the related technical details  we refer the reader to Ref.~\cite{Egner:2024azu}. 
The Fierz symmetry of \eq{eq:fierz-relation-WC} only holds for the contributions of current-current diagrams, but not for penguin diagrams like those in \fig{fig:NLO-Peng-Like}, for which we will not use Fierz-transformed operators.

It is worthwhile to discuss the difference between the result in \eq{eq:ev-db1} and the corresponding 
expression in Eq.~(A.2) of Ref.~\cite{Buras:2006gb}, in which the evanescent operators have been fixed to 
avoid mixing of $Q_+\equiv Q_2+Q_1$ into $Q_-\equiv Q_2-Q_1$ under the renormalisation group evolution, extending the NLO prescription of Ref.~\cite{Buras:1989xd} to NNLO. The non-mixing condition is equivalent to 
the feature that the \emph{renormalised}\ two-loop matrix element $\langle Q_{\pm}\rangle^{(2)}$ is 
proportional to $\langle Q_{\pm}\rangle^{(0)}$, i.e.\ there is no term involving 
$\langle Q_{\mp} \rangle^{(0)}$ when the sum of the two-loop diagrams is expressed in terms of the tree-level matrix elements. 
Since the renormalised matrix element involves counterterm contributions 
with evanescent operators, the latter feature cannot be expected to hold for the \emph{bare}\ matrix 
elements and the counterterm diagrams individually. Now $Q_+$ is even under Fierz transformation 
(in the sense $Q_+^{q_1q_2q_3}\to  Q_+^{q_3q_2q_1}$) while 
$Q_-$ is odd, thus the procedure of Refs.~\cite{Buras:1989xd,Buras:2006gb} maintains 
Fierz symmetry for the renormalised matrix elements in the sense that the Fierz parity  of
$\langle Q_{\pm} (\mu) \rangle $ is equal to $\pm 1$ at all values of the renormalisation scales $\mu$. 

In the calculations of this paper and Ref.~\cite{Egner:2024azu}   Fierz symmetry is instead implemented in a different way. The Fierz transform 
of $Q_{1,2}$ is equal to $Q_{2,1}$ up to an exchange of quark flavours (see \eq{eq:fierz-transformation}), 
which is irrelevant as long as 
no penguin diagrams contribute to $\langle Q_{1,2}\rangle $. The definition in \eq{eq:ev-db1} ensures that
$\langle Q_{1,2}\rangle^{(2)}{}^{\rm bare} = \langle Q_{2,1}\rangle^{(2)}{}^{\rm bare}$  and the equality of the 
renormalised matrix elements in the original and fierzed basis follows from the fact that the step from the 
original to the fierzed basis simply interchanges the evanescent operators associated with $Q_1$ and $Q_2$. 
The latter feature is absent in Ref.~\cite{Buras:2006gb}, because in that paper the CMM basis \cite{Chetyrkin:1997gb} is used, in which the Fierz transformation neither interchanges the two operators
$Q_{1,2}^{\rm CMM}$  nor their corresponding evanescent operators. Our condition that $\langle Q_{1,2}\rangle $ shall agree with the matrix elements of the corresponding fierzed operators is sufficient to ensure the non-mixing of $Q_{\pm}$ into $Q_{\mp}$.

The validity of Eq.~\eqref{eq:fierz-relation-WC} beyond LO, which is justified by the choice of evanescent operators listed above, thus implies
\begin{equation}
  \Gamma^{q_1 q_2 q_3} =
  \widetilde \Gamma^{q_2 q_1 q_3}
  \Big\vert_{\widetilde C_1 \to C_2, \, \widetilde C_2 \to C_1},
  \label{eq::Gam_Gamtil}
\end{equation}
which we use for the remainder of this paper. 

As can be seen from the diagrams in  Figs.~\ref{fig:NLO-Peng-Like} and~\ref{fig:LO-Peng},
there is no need to  use Fierz-arranged operators for the penguin diagrams.
It is possible to treat this class of diagrams differently from the current-current diagrams, 
since it forms a gauge invariant 
 subset.
The general expression of the decay width for this subclass of diagrams is the sum of terms of the form
\begin{align}
   \label{eq:decay-width2}
   \Gamma^{q_1^{(\prime)} q_2^{(\prime)} q_3^{(\prime)}} = \frac1{2M_{H_b}}&\sum_{i,j=1,2} \left(\frac{G_F\;}{\sqrt{2}}\right)^2 \lambda_{q_1q_2q_3}\lambda^*_{q_1^\prime q_2^\prime q_3^\prime}\nonumber\\
   & C_i(\mu_1) C_j (\mu_1) \; \mbox{Im}\, \langle H_b|\, i\int\mbox{d}^4x \;e^{i qx}\;T\left[Q_i^{\dagger q^\prime_1q^\prime_2q^\prime_3}(x)Q_j^{q_1q_2q_3}(0)\right]|H_b\rangle.
\end{align}
Here, the notation on the left-hand side reflects the possibility that there may be an insertion of $\db 1$ effective operators with different quark flavours,
see Fig.~\ref{fig:NLO-Peng-Like}.

As is well-known, the Wilson coefficients $C_i$ that enter the calculation of the decay width mix upon renormalisation. Explicitly,
\begin{equation}
    \label{eq:WC-mixing-matrix}
    C_{i,B} = Z_{ij}C_j(\mu_1), 
\end{equation}
where the index $C_{i,B}$ and $C_i(\mu_1)$ denote the bare and renormalised Wilson coefficients, respectively, and $Z_{ij}$ is the renormalisation matrix, which is obtained by imposing appropriate conditions on the matrix elements of the operators $Q_i$.~In our work, we use the results obtained in the $\msbar$ renormalisation scheme, which are known in the literature to NNLO accuracy~\cite{Buras:1991jm,Buras:1992tc,Ciuchini:1993vr,Chetyrkin:1997gb,Egner:2024azu}. In particular, we use the explicit results given in Appendix A of Ref.~\cite{Egner:2024azu}, which include the mixing with the evanescent operators in Eq.~\eqref{eq:ev-db1}.

\subsection[Expansion in terms of ${\Delta B =0}$ operators]{Expansion in terms of $\mathbf{\Delta B =0}$ operators\label{subs:db0}}

As mentioned above, after using the optical theorem, we perform a HQE and expand the bi-local correlator obtained in increasing powers of $ {\Lambda_{\textrm{QCD}}}/{m_b}$
\begin{equation}
    \label{eq:HQE-transition-operator}
    \mathcal{T}=\left[ \mathcal T_0 + \mathcal T_2 +\mathcal T_3 \right]\left[1+\mathcal{O}(\Lambda_{\textrm{QCD}}^4/m_b^4)\right],
\end{equation}
where $\mathcal T_n$ denotes the part of the decay width that is suppressed by a factor of $(\Lambda_{\textrm{QCD}}/m_b)^n$. 
The leading terms in this expansion, namely $\mathcal{T}_0$ and $\mathcal{T}_2$, are given by the free $b$ quark decay and the chromomagnetic interactions of the final state quarks with the hadronic cloud surrounding the $b$ quark, respectively. 
$\langle H_b | \mathcal T_0 | H_b \rangle$ is  known to NNLO accuracy~\cite{Egner:2024azu,Egner:2024lay}, but contributes equally to all decay rates and drops out from lifetime differences.
The contributions from $\langle H_b | \mathcal T_2 | H_b 
\rangle$ have negligible effects on the difference in lifetimes 
of $B^+$ and $B_d^0$, due to the isospin symmetry of QCD. 

\begin{figure}[t]
    \begin{minipage}{.5\textwidth}
        \centering
        \includegraphics[width=1\linewidth]{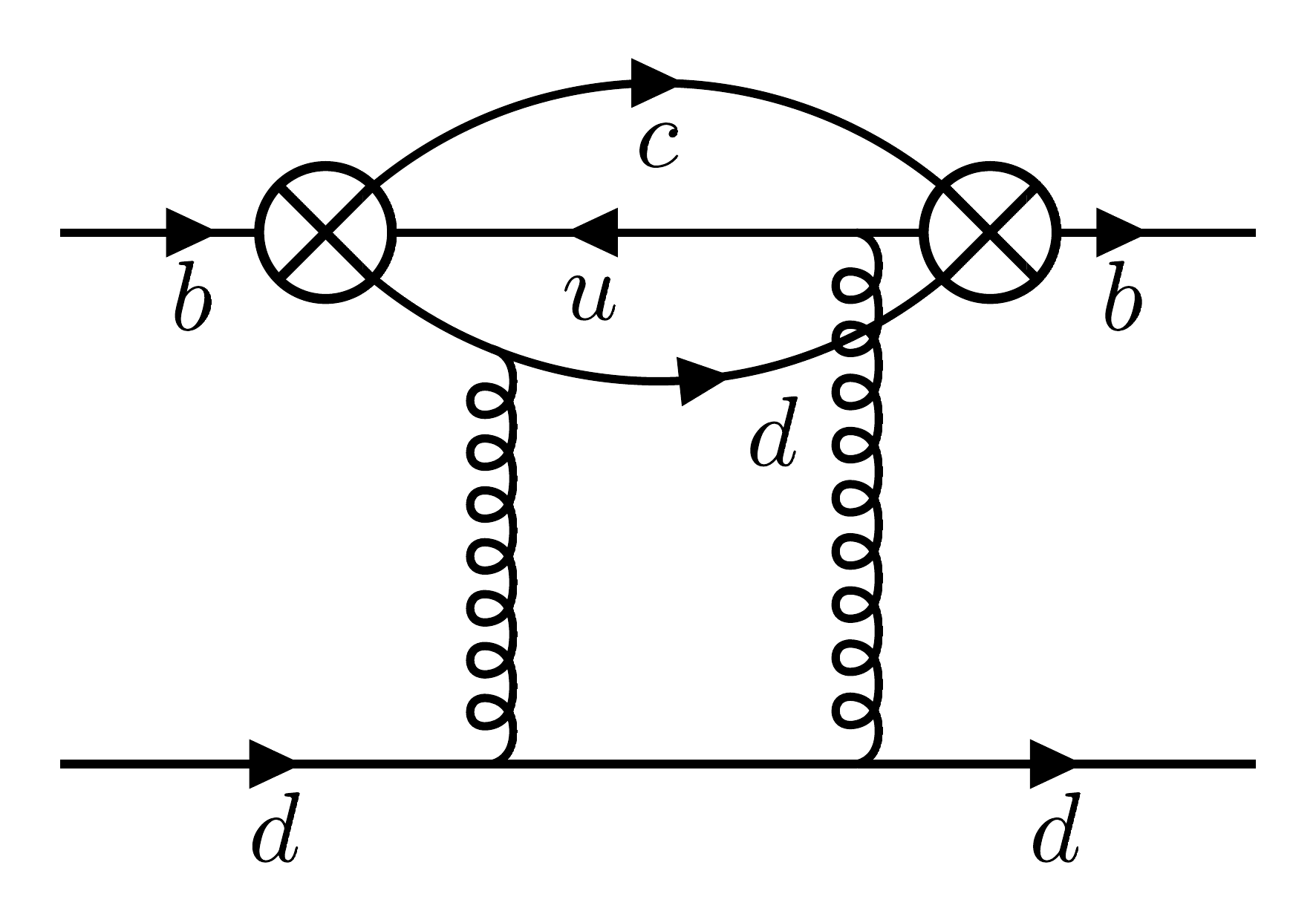}
    \end{minipage}%
    \begin{minipage}{.5\textwidth}
        \centering
        \includegraphics[width=0.8\linewidth]{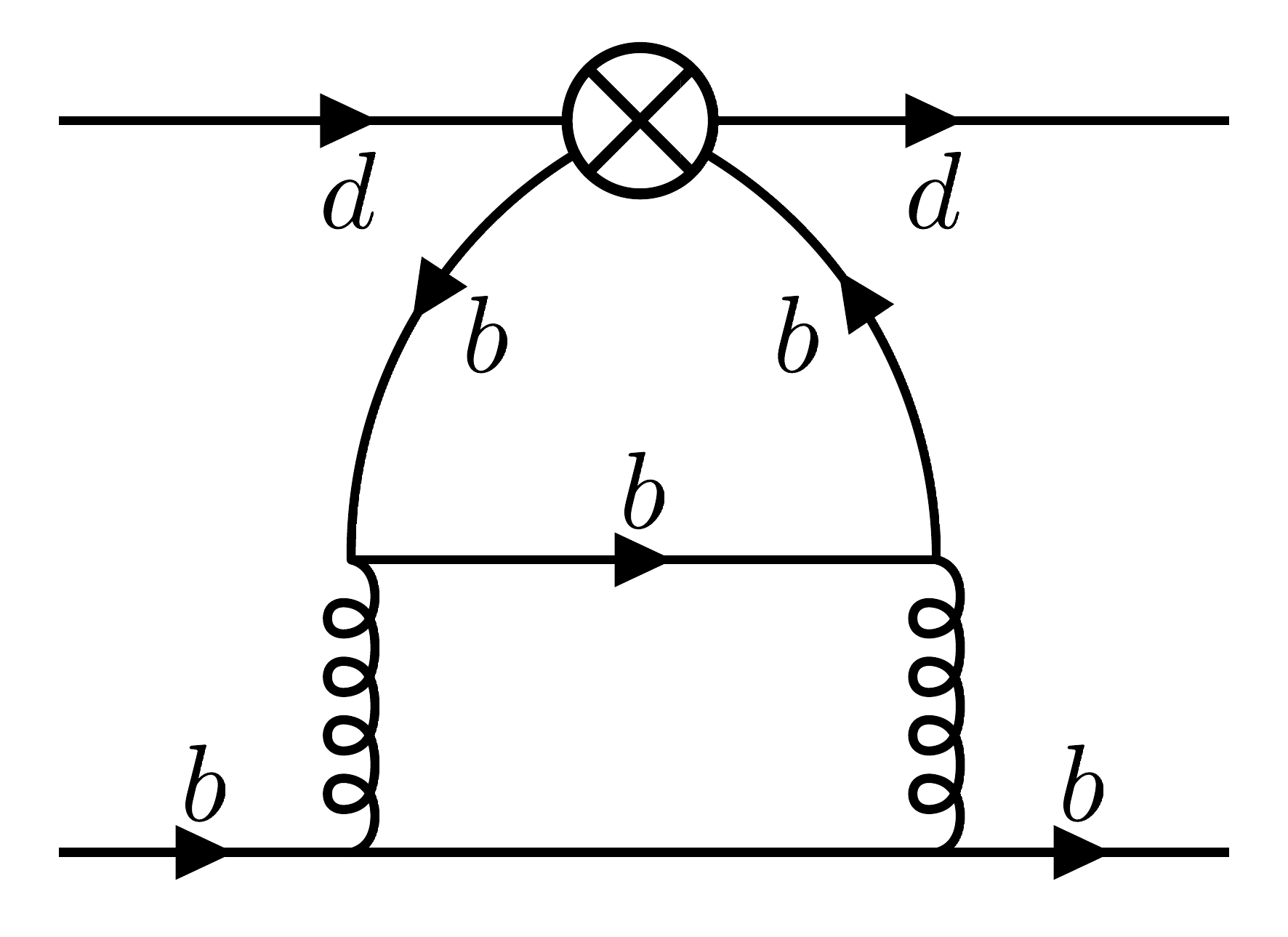}
    \end{minipage}%
    \caption{
    Example of isospin singlet contributions to the decay rate in the $|\Delta B|=1$ (left) and $\Delta B = 0$ theory (right). These are often referred to as \textit{penguin} or \textit{eye contraction} diagrams. They cancel in the lifetime ratio $\tau(B^+)/\tau(B_d^0)$.}
    \label{fig:peng-contr}
\end{figure}

At $\mathcal{O}\left(\Lambda^3_{\rm QCD}/m_b^3\right)$ we encounter the first corrections to the lifetimes difference. These come from weak interactions between the $b$ quark and the light valence quarks and can be written in the form
\begin{equation}
    \mathcal{T}_3 = \mathcal{T}^{u} + \mathcal{T}^{d} +\mathcal{T}_{sing},
\end{equation}
where the superscript $u$ refers to the pairs of up-type quarks inside the loop diagram in Fig.~\ref{fig:wa-pi-LO}, while $d$ indicates the internal $ud$ and $cd$ pairs. 
$\mathcal{T}_{sing}$ refers to additional isospin singlet corrections, which describe power-suppressed contributions to the free quark decay from strong interactions with the spectator quark (see diagrams in Fig.~\ref{fig:peng-contr}). These cancel in the difference $\Gamma(B^+)-\Gamma(B_d^0)$.

Contributions to $\mathcal{T}^{u}$ and $\mathcal{T}^{d}$ involve the following dimension-6 operators
\begin{eqnarray}
    Q^q &=& (\overline{b}\gamma_\mu P_L q) \; (\overline{q} \gamma^\mu P_L b)\nonumber\\
    Q_S^q &=& (\overline{b} P_L q) \; (\overline{q} P_R b) \nonumber\\
    T^q &=& (\overline{b}\gamma_\mu P_L T^a q )\; (\overline{q} \gamma^\mu P_L T^a b) \nonumber\\
    T_S^q &=& (\overline{b}P_L T^a q) \; (\overline{q} P_R T^a b)\,.\label{eq:eff-db0-ham}
\end{eqnarray}

In addition to the physical operators, we define the first-generation evanescent operators as
\begin{eqnarray}
    & E[Q^q] &= (\overline{b}\gamma^{\mu_1} \cdots\gamma^{\mu_3}P_L q)\; (\overline{q}\gamma_{\mu_3} \cdots\gamma_{\mu_1}P_L b) - (4-8\epsilon+a\epsilon^2)Q^q,\nonumber\\
    & E[Q^q_S] & = (\overline{b}\gamma^{\mu_1} \gamma^{\mu_2} P_L q)\; (\overline{q}\gamma_{\mu_2} \gamma_{\mu_1} P_R b ) - (4-8\epsilon+a_S\epsilon^2)Q^q_S,\label{eq:1st-gen-ev-db0}
\end{eqnarray}
the second-generation evanescent operators as
\begin{eqnarray}
    & E[Q^q]^{(2)} & = (\overline{b}\gamma^{\mu_1}\cdots \gamma^{\mu_5} P_L q)\; (\overline{q}\gamma_{\mu_5}\cdots \gamma_{\mu_1} P_L b) - (16+b\epsilon+c\epsilon^2)Q^q,\nonumber\\
    & E[Q^q_S]^{(2)} & = (\overline{b}\gamma^{\mu_1}\cdots \gamma^{\mu_4} P_L q)\; (\overline{q}\gamma_{\mu_4}\cdots \gamma_{\mu_1} P_R b) - (16+b_S\epsilon+c_S\epsilon^2)Q^q_S,\label{eq:2nd-gen-ev-db0}
\end{eqnarray}
and finally, the third-generations evanescent operators as
\begin{eqnarray}
    & E[Q^q]^{(3)} & = (\overline{b}\gamma^{\mu_1} \cdots\gamma^{\mu_7} P_L q)\; (\overline{q}\gamma_{\mu_7}\cdots \gamma_{\mu_1} P_L b) - (64+e\epsilon+f\epsilon^2)Q^q,\nonumber\\
    & E[Q^q_S]^{(3)} & = (\overline{b}\gamma^{\mu_1} \cdots\gamma^{\mu_6} P_L q)\; (\overline{q}\gamma_{\mu_6} \cdots\gamma_{\mu_1} P_R b) - (64+e_S\epsilon+f_S\epsilon^2)Q^q_S.\label{eq:3rd-gen-ev-db0}
\end{eqnarray}
In our work, we choose the same definitions of evanescent operators related to physical operators 
with the same Dirac (but different colour) structure, meaning that 
$E[T^q]^{(i)}$ and $E[T_S^q]^{(i)}$ also involve the coefficients of the $\epsilon$ and $\epsilon^2$ terms appearing in Eqs.~\eqref{eq:1st-gen-ev-db0} to~\eqref{eq:3rd-gen-ev-db0}.
Unlike in the case of $\db1$ operators, 
there is no preferred choice for the terms of $\mathcal{O}(\epsilon)$ and 
$\mathcal{O}(\epsilon^2)$, we only fix the value of 8$\epsilon$ in Eq.~\eqref{eq:1st-gen-ev-db0} to facilitate the comparison with the 
literature, see e.g.~Ref.~\cite{Beneke:2002rj}.
In the calculation of lifetimes differences among mesons there is no need to maintain Fierz symmetry 
in the definition of the coefficients of the $\Delta B=0$ operators, because the quark fields in \eq{eq:eff-db0-ham} are already arranged in the correct way for the computation of the hadronic matrix elements.
Any dependence on the coefficients
$a,\ldots,f_S$ will cancel
as long as the 
same evanescent operators
are used in the calculation of the matrix elements. Specifically, in the gradient-flow method this dependence cancels in the matching to the flowed operators, which are subsequently matched to the operators of lattice QCD
\cite{Black:2023vju,Black:2024iwb,Black:2026dzp,Black:2026rbz}. 
For simplicity, in our analysis we take the coefficients $a,\ldots,f_S = 0$.

The situation is different for the calculation of lifetime differences of baryons which involves only 
two of the four operators in \eq{eq:eff-db0-ham} at dimension-6 level, while two linear combinations of these operators have power-suppressed baryonic matrix elements after a Fierz transformation~\cite{Neubert:1996we,Beneke:2002rj}. Thus in applications to baryons Fierz symmetry matters.

When we perform the matching between the ``full" $\db 1$ side and the effective $\db 0$ theory, the weak interactions with the valence quarks take the following expression
\begin{align}
    \mathcal{T}^{u}= \frac{G_F^2 m_b^2}{6\pi}
        &\left[
            |\lambda_{cud}|^2\left( F^{cu} Q^d + F_S^{cu} Q^d_S + G^{cu} T^d+ G_S^{cu} T_S^d\right)\right.\nonumber\\
            &\left. \; + |\lambda_{ccd}|^2\left( F^{cc} Q^d + F_S^{cc} Q^d_S + G^{cc} T^d+ G_S^{cc} T_S^d\right) \nonumber \right.\\
            &\left.  \; + |\lambda_{uud}|^2\left( F^{uu} Q^d + F_S^{uu} Q^d_S + G^{uu} T^d+ G_S^{uu} T_S^d\right) \right.\nonumber\\
            &\left. \;+\lambda_{cud}^\prime \left( {\tilde F^{cu}} Q^d + \tilde F_S^{cu} Q^d_S + \tilde G^{cu} T^d+ \tilde G_S^{cu} T_S^d\right) 
            \right],\nonumber\\
     \mathcal{T}^{d} = \frac{G_F^2 m_b^2}{6\pi}&\left[|V_{cb}|^2\left(F^{cd} Q^u + F_S^{cd} Q_S^u + G^{cd} T^u + G_S^{cd} T_S^u\right) \right .\nonumber\\
     &\left. + |V_{ub}|^2 \left(F^{ud} Q^u + F_S^{ud} Q_S^u + G^{ud} T^u + G_S^{ud} T_S^u\right.\right.\nonumber\\
     &\left.\left. \hspace{1.5cm}
     {+ F^{cs} Q^u + F_S^{cs} Q_S^u + G^{cs} T^u + G_S^{cs} T_S^u} \right) \right]\label{eq:tu-td},
\end{align}
where the superscripts in the Wilson coefficients $F_{(S)}^{qq^\prime}$ and $G_{(S)}^{qq^\prime}$ refer to the $qq^\prime$ pair in the loop, while the superscripts on the dimension-6 operators refer to the external light spectator quarks, see Fig.~\ref{fig:wa-pi-LO}. 
The contributions $\tilde F^{cu},\ldots,\tilde G_S^{cu}$ come from the CKM-suppressed one-particle-reducible diagrams as in Fig.~\ref{fig:NLO-Peng-Like} where both a $uu$ and a $cc$ pair are considered.

$\mathcal{T}^{u}$ in \eq{eq:tu-td} is easily generalised to the case in which the valence quark is $s$ instead of $d$, with the appropriate replacement in CKM factors and operators. However, this replacement also 
interchanges CKM-favoured with CKM-suppressed terms and one should keep in mind that we only quote the CKM-favoured parts of $\mathcal{T}^{u}$ at NNLO.

\begin{figure}[t]
\begin{center}
    \begin{minipage}{.5\textwidth}
        \centering
        \includegraphics[width=1\linewidth]{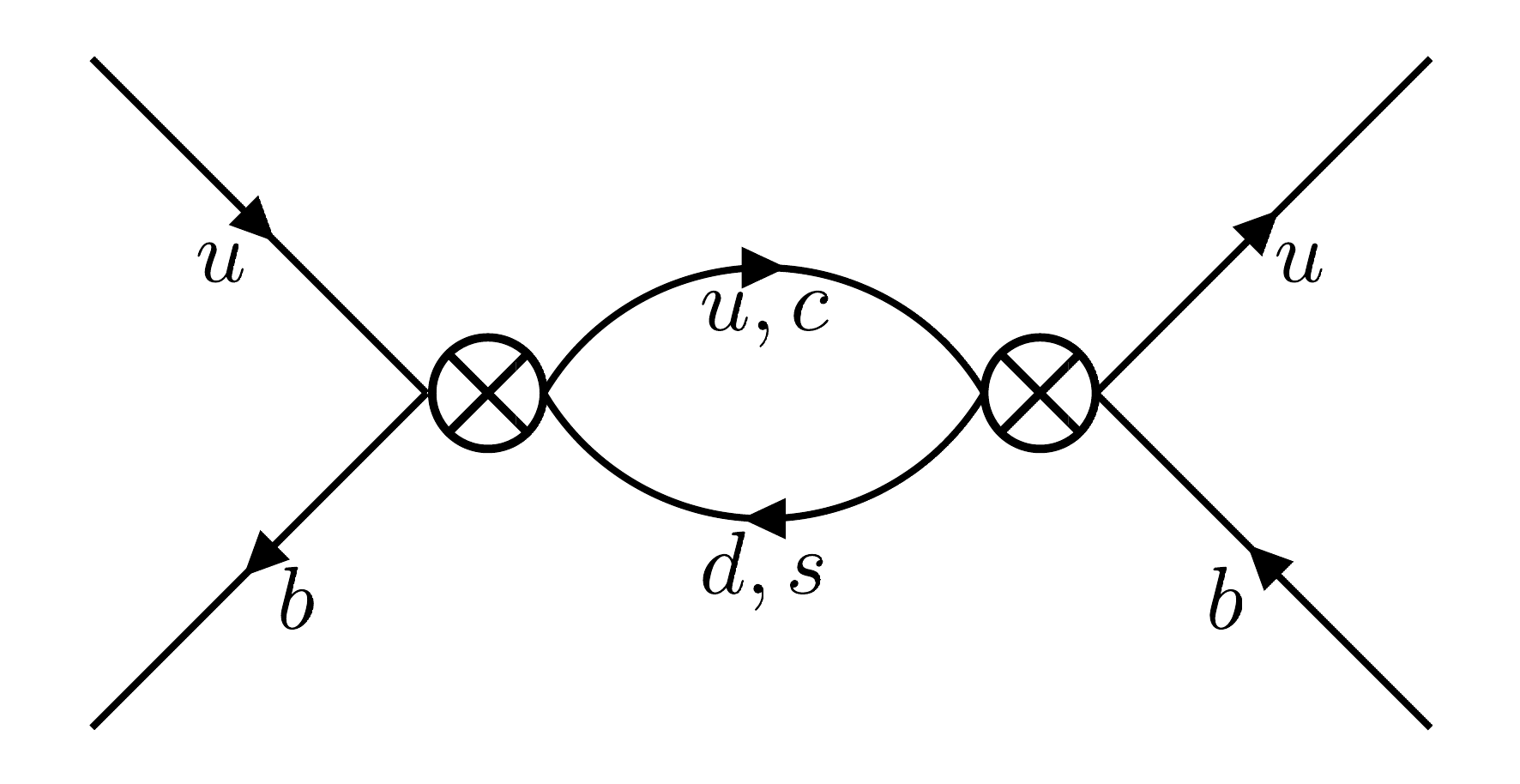}
    \end{minipage}%
\end{center}
    \caption{Sample weak annihilation-type Feynman diagrams contribution to the LO decay rate
    of a $B^+$ meson. 
}
    \label{fig:LO_GamBplus_WA}
\end{figure}

For the derivation of ${\cal T}^{d}$, we have used $|V_{ud}|^2 + |V_{us}|^2 \approx 1$, $|V_{cs}|^2 + |V_{cd}|^2 \approx 1$ and $m_d=m_s=0$, so that $F^{qs}_{(S)}=F^{qd}_{(S)}
$ and $G^{qs}_{(S)}=G^{qd}_{(S)}$. Note that
the terms  proportional to $|V_{ub}|^2$ get contributions from WA and PI. The latter are obtained from the $cd$ terms by taking the limit $m_c\to0$.
On the other hand, the WA contributions to $\Gamma(B^+)$ originate from Feynman diagrams as shown in Fig.~\ref{fig:LO_GamBplus_WA}.
They contribute to all $|V_{ub}|^2$-suppressed
terms of ${\cal T}^{d}$ and are conveniently obtained from the $cu$ contributions to ${\cal T}^{u}$. This is possible thanks to our choice of evanescent operators in Eq.~\eqref{eq:ev-db1} which preserves Fierz symmetry, but we stress that in extracting these terms from the $cu$ contributions, we interchange the role of the $\db 1$ Wilson coefficients. Explicitly, we have
\begin{equation}
    \label{eq:WC-fierzed-Bplus}    \left(F^{cs},F_S^{cs},G^{cs},G_S^{cs}\right)=\left(F^{cu},F_S^{cu},G^{cu},G_S^{cu}\right)|_{C_1\leftrightarrow C_2},
\end{equation}
with analogous expressions for $F^{ud}$ (where the additional limit $m_c\to 0$ is taken).
We include them up to NLO.

We define the operators in the $\db 0$ basis at the renormalisation scale $\mu_0$, which we later take to be $\mu_0=\overline m_b(\overline m_b)$~\cite{Black:2024bus}. The dependence of $\mu_0$ in the corresponding matrix elements cancels against the $\mu_0$ dependence of the Wilson coefficients $F_{(S)}^{qq^\prime}$ and $G_{(S)}^{qq^\prime}$. These coefficients depend also on the $\db 1$ renormalisation scale $\mu_1$, due to the truncation of the perturbative series, and on the ratio $x=m_c/m_b$. The residual dependence on $\mu_1$ is expected to diminish with increasing order of $\alpha_s$.
For this reason it is often used as an estimate of the uncertainty
form higher order pertubative corrections, as discussed later in Section~\ref{sec::phen}. 

As can be seen from Eq.~\eqref{eq:tu-td}, and as we mentioned earlier, the contributions to the decay rate that contain $cc$ or $uu$ intermediate states, e.g.~$F^{cc}_{(S)}$ and $G_{(S)}^{cc}$, are CKM-suppressed with respect to the contributions coming from $cu$ pairs. In Ref.~\cite{Beneke:2002rj}, the NLO corrections to $\mathcal{T}^{u}$ have been computed in the approximation $|V_{ub}|=0$, $|V_{cd}|=0$ and $|V_{ud}|=1$, thus neglecting the CKM-suppressed contributions, whose impact was estimated to be at the sub-percent level. 
We compute $F^{cc}_{(S)}$, $G_{(S)}^{cc}$, $F^{uu}_{(S)}$, $G_{(S)}^{uu}$, $\tilde F^{cu}_{(S)}$ and $\tilde G_{(S)}^{cu}$ to NLO accuracy. In addition, we include the CKM-suppressed contrubutions to $\mathcal{T}^{d}$, that is $F^{ud},\ldots,G_S^{cs}$ to NLO accuracy. We confirm that their contribution is well below the percent level, see Section~\ref{sec::phen}. Thus, we extend the approximation of Ref.~\cite{Beneke:2002rj} to NNLO.
The terms which are neglected  introduce an uncertainty of the order of $|V_{cd}|^2 \alpha_s^2 m_c/m_b$, which is below 1\textperthousand.

The final step to obtain a theoretical prediction for the lifetimes is to compute the
hadronic matrix elements of the effective operators appearing in Eq.~\eqref{eq:eff-db0-ham}. They enter the calculation in the isospin breaking contributions and can be parametrised according to~\cite{Neubert:1996we,Beneke:1996gn}
\begin{align}
    &\langle B^+| \left(Q^u - Q^d\right)(\mu_0)| B^+ \rangle = f_B^2 M_B^2 B_1(\mu_0), \nonumber\\
    &\langle B^+| \left(Q^u_S - Q^d_S\right)(\mu_0)| B^+ \rangle = f_B^2 M_B^2 B_2(\mu_0)\nonumber\\
    &\langle B^+| \left(T^u - T^d\right)(\mu_0)| B^+ \rangle = f_B^2 M_B^2 \epsilon_1(\mu_0), \nonumber\\
    &\langle B^+| \left(T^u_S - T^d_S\right)(\mu_0)| B^+ \rangle = f_B^2 M_B^2 \epsilon_2(\mu_0).\label{eq:ome} 
\end{align}
Here, $f_B$ is the decay constant of the B meson. $B_{1,2}$ and $\epsilon_{1,2}$ are referred to as \textit{bag parameters} and need to be determined using non-perturbative methods, namely lattice QCD (LQCD)~\cite{DiPierro:1998ty,DiPierro:1999tb,Becirevic:2001fy,Lin:2022fun,Black:2023vju,Black:2024iwb} and heavy quark effective theory (HQET) sum rules~\cite{Kirk:2017juj,King:2021jsq,Black:2024bus}.

Using the isospin relation $\langle B^+|\{Q,T\}^{u,d}|B^+\rangle = \langle B_d^0|\{Q,T\}^{d,u}|B_d^0\rangle$, we can then write the difference of the decay widths $\Gamma(B_d^0)$ and $\Gamma(B^+)$ as
\begin{align}
    \Gamma(B_d^0) - \Gamma(B^+) = \frac{G_F^2 m_b^2}{12\pi}f_B^2 M_B &\left[ |\lambda_{cud}|^2 \vec{F}^{cu}+|\lambda_{ccd}|^2\vec{F}^{cc}+|\lambda_{uud}|^2\vec{F}^{uu}\right.\nonumber\\
    &\left.\;\; +\lambda_{cud}^\prime {\vec{\tilde F}^{cu}}- |V_{cb}|^2 \vec{F}^{cd} - |V_{ub}|^2\left(\vec{F}^{ud}{+\vec{F}^{cs}}\right)\right]\cdot \vec{B},
    \label{eq::diff}
\end{align}
where we have introduced the shorthand notation 
\begin{eqnarray}
    \vec{F}^{qq^\prime} = \begin{pmatrix}
        F^{qq^\prime} (\mu_1,\mu_0,x) \\
        F_S^{qq^\prime}(\mu_1,\mu_0,x)\\
        G^{qq^\prime}(\mu_1,\mu_0,x)\\
        G_S^{qq^\prime}(\mu_1,\mu_0,x)
    \end{pmatrix}, &&
    \vec{B} = \begin{pmatrix}
        B_1 (\mu_0)\\
        B_2(\mu_0)\\
        \epsilon_1(\mu_0)\\
        \epsilon_2(\mu_0)
    \end{pmatrix}.
    \label{eq::F_B}
\end{eqnarray}

In our analysis of $\tau(B^+)/\tau(B_d)$, we use results for the bag parameters presented in Ref.~\cite{Black:2024bus}, which have been derived from HQET sum rules. 

\subsection[${\Delta B = 0}$ renormalisation constants]{\label{sub::rendb0}$\mathbf{\Delta B = 0}$ renormalisation constants}





The Wilson coefficients $\vec F^q$ given in Eq.~\eqref{eq::F_B} mix upon renormalisation and follow a similar relation as in Eq.~\eqref{eq:WC-mixing-matrix}.
Unlike for the renormalisation matrix $\hat Z$ needed in the $\db 1$ part of our computations, there are, to our knowledge, no public results of the $\db 0$ renormalisation constants in a form that can be used in our work. Recently, a comprehensive study has been published~\cite{Aebischer:2025hsx} where the $\hat{Z}$ matrix for the $\db 0$ 
operator basis, 
and many others, is derived. 
Yet, in this work, the results are obtained with a fixed scheme for the evanescent operators (see Appendix~A of Ref.~\cite{Aebischer:2025hsx}). As stated earlier in Section~\ref{subs:db0}, we want to keep the definition of the evanescent operators general. For this reason, we have computed the renormalisation constants ourselves. We refrain from presenting explicit results for these quantities here, but refer to the supplementary material which can be downloaded from~\cite{progdata}, where the $\hat Z$ matrix can be found in 
computer-readable format.

We have cross-checked 
our results for the $Z$ factors by performing two independent calculations  within our collaboration. Furthermore,  we have performed the computation in general $R_\xi$ gauge of QCD, and observe that the gauge parameter $\xi$ drops out from the final results. We have also compared our results with an independent calculation performed in Ref.~\cite{Black:2026dzp,Black:2026rbz}
and we have found complete agreement.




\section{Calculation}

In our work, we employ dimensional regularisation with anticommuting $\gamma_5$. 
For the calculation of the $\db 1$ and $\db 0$ amplitudes up to three and two loops, respectively, we have used two independent computer codes developed in \texttt{FORM}~\cite{Kuipers:2012rf,Davies:2026cci} and \texttt{Mathematica}. 

The starting point of our routine is the generation of Feynman diagrams. This is done using \texttt{qgraf}~\cite{Nogueira:2006pq}. 
As mentioned in Section~\ref{sec::eft}, we neglect isospin singlet contributions
(see Fig.~\ref{fig:peng-contr}) to the total decay rate. 
At NNLO, in addition, we consider the CKM-favoured processes only, neglecting the CKM-suppressed terms proportional to 
$|\lambda_{ccd}|^2$, $\lambda^\prime_{cud}$ and $|\lambda_{uud}|^2$.
Effectively, at this order, we set $|V_{ub}|=0$,
$|V_{cd}|=0$ and $|V_{ud}|=1$.

We process the Feynman amplitudes using \texttt{tapir}~\cite{Gerlach:2022qnc}, which outputs them as \texttt{FORM} code. This step requires some attention, since we deal with four-fermion effective vertices. 
We find it convenient to separate the four-particle vertices in two three-particle vertices, introducing an auxiliary particle. The corresponding Feynman rules
are chosen such that the Lorentz and colour structure of the desired operators is reproduced correctly. This procedure overcomes any possible relative sign problem between different diagrams that could arise in the reconstruction of the spinor chain. 

The Feynman amplitudes obtained are then processed using an in-house setup called \texttt{calc}, 
in which we implement a routine that expresses the tensor integrals that appear in the calculations in terms of scalar Feynman integrals via suitable projectors. This routine is detailed in Ref.~\cite{Reeck:2024iwk}. 
After this step, the amplitudes are expressed in terms of a basis of spinor structures, given in Appendix~\ref{app::basis}, and corresponding coefficients that depend only on the dimension $d$, internal masses, and scalar products of momenta.

%
Afterwards, we make use of \texttt{Kira}
\cite{Maierhofer:2017gsa,Klappert:2020nbg,Lange:2025ofh} to perform an
integration-by-parts (IBP) reduction to master integrals.
We find that roughly 70\% of all the scalar integrals that appear are present already in the IBP tables derived in the computation of the mixing in the neutral $B$ systems~\cite{Reeck:2024iwk,Gerlach:2025tcx}, while we performed a complementary reduction to cover the remaining 30\%. As expected, no new master integrals appear, and we implement the results detailed in the former works. The master integrals are given in terms of a semi-analytic log-polynomial series in $x=m_c/m_b$; explicit expressions can be found  in the ancillary files of Ref.~\cite{Reeck:2024iwk}. 

Finally, we perform the renormalisation of the $\db{1}$ and $\db{0}$ side separately.
We renormalise $\alpha_s$, the charm quark mass
and the matching coefficients of the effective operators
in the $\overline{\rm MS}$ scheme and the bottom
quark mass in the pole scheme. The latter is
transformed afterwards to more appropriate renormalisation schemes; see Section~\ref{sec::phen} for more details.
We refrain from renormalising the wave function of the external quarks since the corresponding $Z$ factor
drops out in the matching procedure.
Afterwards, we perform the matching to extract the Wilson coefficients $\vec F^{qq^\prime}$, which is described in the next Section.




\section{Matching coefficients to NNLO}

In Sections~\ref{subs:db1} and~\ref{subs:db0} we have described the calculation of the
amplitudes in the $|\Delta B|=1$ and $\Delta B =0$ theories. They are
ultraviolet finite but still contain infrared poles in $\epsilon$ beyond
LO. This requires that the matching of the two theories has to be performed to
higher order in $\epsilon$. For a further discussion we refer
to Section~2.7 of Ref.~\cite{Gerlach:2025tcx} where the matching procedure
between $|\Delta B|=1$ and $|\Delta B|=2$ theories is described is detail
including examples demonstrating the role of evanescent operators.

We define the perturbative expansion of the matching coefficients  in
Eq.~(\ref{eq:tu-td}) as follows
\begin{eqnarray}
  F &=& F^{(0)} + \frac{\alpha_s}{4\pi} F^{(1)} 
        + \left( \frac{\alpha_s}{4\pi} \right)^2 F^{(2)}
        + {\cal O}(\alpha_s^3)
\end{eqnarray}
where $\alpha_s\equiv \alpha_s(\mu_1)$.  We refrain from presenting explicit
results for the matching coefficients but refer the supplementary material
which contain results for $F^{(0)}$, $F^{(1)}$ and $F^{(2)}$~\cite{progdata}.

\begin{figure}[t]
\centering
    \begin{tabular}{ccc}
         \includegraphics[height=0.3\linewidth]{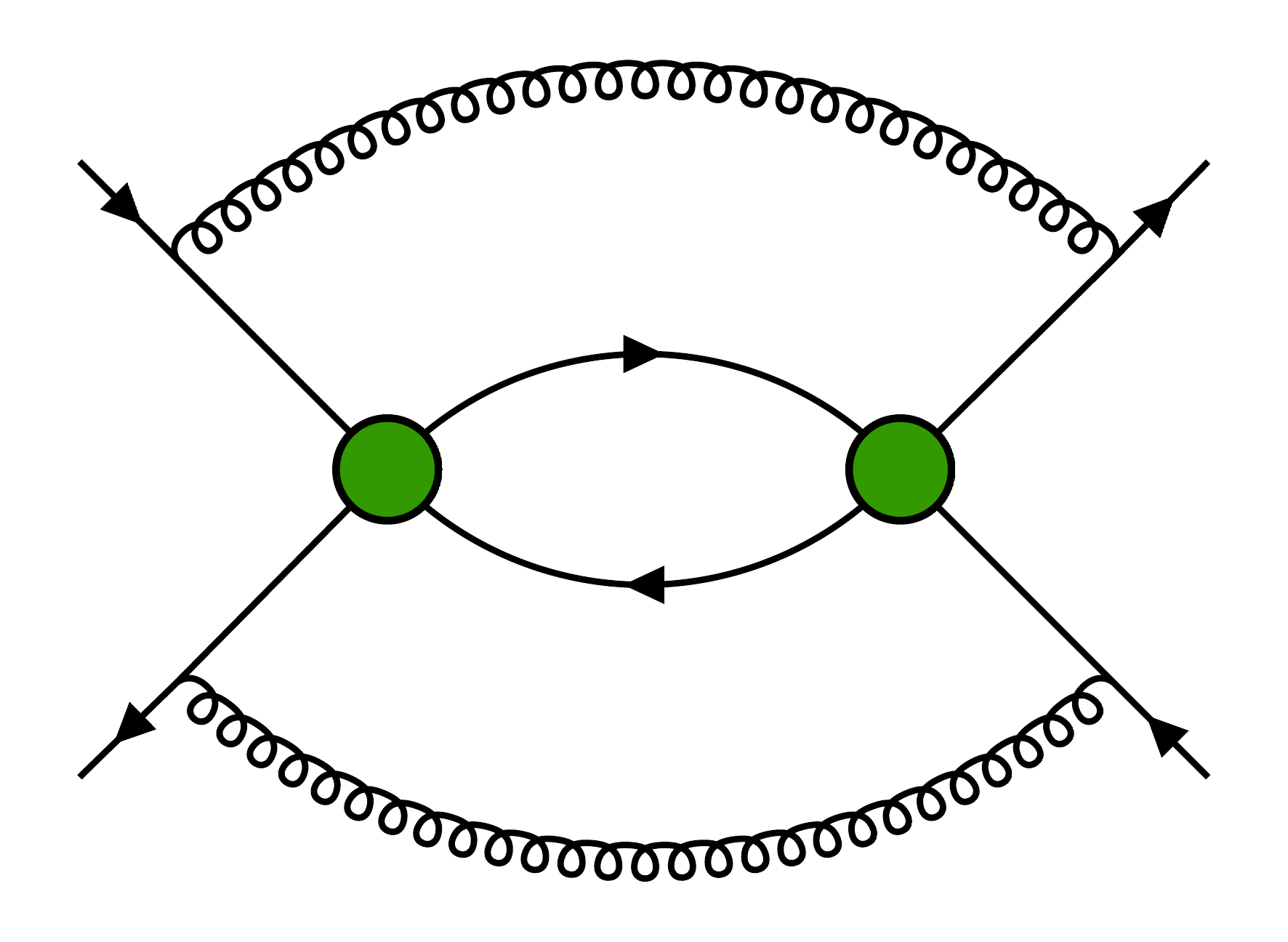} & \hspace{.6cm}\raisebox{2cm}{\begin{Huge}\boldmath$\Longleftrightarrow$\end{Huge}}& \hspace{.8cm}\includegraphics[height=0.28\linewidth]{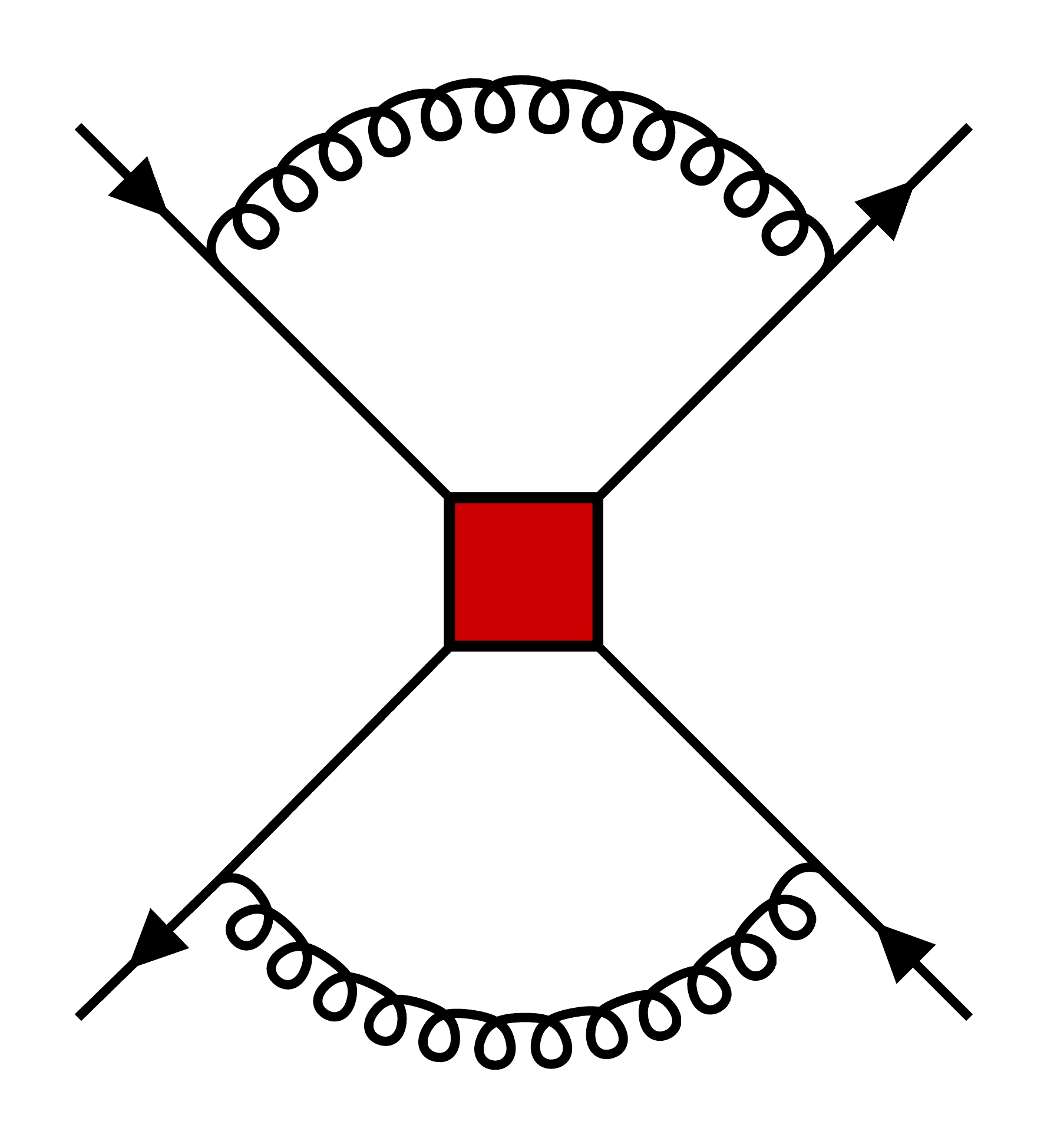}  
    \end{tabular}
    \caption{
    Example of selected diagrams used as a check of correct IR singularities cancellation. A green blob and a red square correspond to the insertion of $|\Delta B|=1$ and $\Delta B =0$ operators, respectively. The diagram on the left is IR divergent, but UV finite, while the diagram on the right contains both IR and UV poles. The latter are canceled by appropriate counterterms, while the former cancel in the matching. }
    \label{fig::diag_IR}
\end{figure}

There are several checks which are fulfilled by our results:
\begin{itemize}
\item
  In the matching procedure all infrared poles cancel
  leading to finite matching coefficients.

  The cancellation of the poles can even be observed for individual diagrams:
  We have identified typical three-loops diagrams at the $|\Delta B|=1$ side
  and the corresponding two-loop diagrams at the $|\Delta B|=0$ side (see
  Fig.~\ref{fig::diag_IR} for examples)  and have
  checked that the difference is finite for $\epsilon\to 0$.

\item We have performed the calculation of all parts for general QCD gauge
  parameter $\xi$ and have checked that in the final result for the matching
  coefficients $\xi$ drops out.  Note that the individual renormalised
  $|\Delta B|=0$ and $|\Delta B|=1$ amplitudes still depend on $\xi$.

  \item The calculations performed in Sections~\ref{subs:db1} and~\ref{subs:db0} assume
  a common renormalisation scale $\mu$ which is present both in the matching
  coefficients $F^{qq^\prime}$, $F_S^{qq^\prime}$, $G^{qq^\prime}$ and $G_S^{qq^\prime}$
  and in the hadronic matrix elements. We use the
  anomalous dimensions of the $|\Delta B|=0$ theory (see Section~\ref{sub::rendb0}) in
  order to separate the scales associated to the two theories, $\mu_0$ and
  $\mu_1$. This leads to the form of the matching coefficients used in
  Eq.~(\ref{eq::F_B}). At this point we can use the anomalous dimensions of
  the $|\Delta B|=1$ theory to cross-check the $\mu_1$ dependence. 

\item We have performed a number of internal cross checks. In particular,
  we have performed the computation of the bare $|\Delta B|=0$ and $|\Delta
  B|=1$ amplitudes, the renormalisation in both theories, and the matching 
  twice using independent computer codes.

\end{itemize}

\section{\label{sec::phen}Phenomenology for $\mathbf{B}$ mesons}

\begin{table}[tbh]
  \begin{center}
  {\scalefont{0.9}
    \renewcommand{\arraystretch}{1.2}
    \begin{tabular}{rcll | rcll}
      \hline
      $\alpha_s(M_Z)$ &=& $\SI{0.1180(9)}{}$ & \cite{ParticleDataGroup:2024cfk}
      &
      $G_F$ &=& $\SI{1.1663787e-5}{GeV^{-2}}$ &  \cite{ParticleDataGroup:2024cfk} 
      \\
      $M_W$ &=& $\SI{80.3629(133)}{GeV}$ &  \cite{ParticleDataGroup:2024cfk}
      &
      $M_Z$ &=& $\SI{91.1880(20)}{GeV}$ &  \cite{ParticleDataGroup:2024cfk}
      \\
      $m_t^{\text{OS}}$ &=& $\SI{172.4(7)}{GeV}$ & \cite{ParticleDataGroup:2024cfk} 
      &                                                 
      $\overline{m}_b(\overline{m}_b)$ &=& $\SI{4.163(16)}{GeV}$ & \cite{Chetyrkin:2010ic} 
      \\
      $\overline{m}_c(\SI{3}{GeV})$ &=& $\SI{0.993(8)}{GeV}$ & \cite{Chetyrkin:2017lif}
      &
      $M_{B}$ &=& $\SI{5279.41(07)}{MeV}$ & \cite{ParticleDataGroup:2024cfk} 
      \\
      $f_{B}$ &=& $\SI{0.1900(13)}{GeV}$ & \cite{FLAG:2024oxs} 
      &
      $\tau(B^+)$ &=& $\SI{1.638(4)}{\pico\second}$ & \cite{ParticleDataGroup:2024cfk}\\
      $|V_{ud}|$ &=& $0.974343^{+0.000049}_{\scalebox{2}[1]{-}  0.000048}$ &
                                                                             \cite{CKMfitter}
      &
      
      $\tau(B^+)/\tau(B)$ &=& $1.076(4)$ & \cite{HFLAV:2024ctg}\\
      
      $|V_{cb}|$ &=& $0.04159^{+0.00023}_{\scalebox{2}[1]{-}  0.00084}$ &
                                                                             \cite{CKMfitter}
       &
      $|V_{cd}|$ &=& $0.22491^{+0.00020}_{\scalebox{2}[1]{-}  0.00022}$ &
                                                                             \cite{CKMfitter}                                                                 
      \\
      $|V_{ub}|$ &=& $0.003743^{+0.000053}_{\scalebox{2}[1]{-}  0.000057}$ &
                                                                             \cite{CKMfitter}
       &
      $\bar{\rho}$ &=& $	0.1562^{+0.0102}_{\scalebox{2}[1]{-} 0.0045}$ &
                                                                             \cite{CKMfitter}\\
      \hline
    \end{tabular}
    }
  \end{center}
  \caption{\label{tab::input}Input parameters for the numerical analysis.  The
    quoted $m_t^{\rm pole}$ corresponds to     $\overline{m}_t(\overline{m}_t)=(162.6 \pm 0.7)$~GeV in the $\overline{\rm MS}$ scheme. }
\end{table}

In this Section we discuss the numerical effect of the new NNLO corrections on
the lifetime ratio $\tau(B^+)/\tau(B_0)$.  We use Eq.~(\ref{eq::diff})
together with the input parameters given in Tab.~\ref{tab::input} and the bag
parameters of the $B$ mesons obtained from QCD sum rules in
Ref.~\cite{Black:2024bus},
\begin{align}
  B_1 &= 1.013^{+0.066}_{\scalebox{2}[1]{-} 0.059}\,,\nonumber\\
  B_2 &= 1.004^{+0.085}_{- 0.081}\,, \nonumber\\
  \epsilon_1 &= -0.098^{+0.026}_{- 0.032}\,,\nonumber\\
  \epsilon_2 &= -0.037^{+0.019}_{- 0.020}\,.
\label{eq:bag_decay_input}
\end{align}
They are defined at the scale $\mu_0=\overline{m}_b(\overline{m}_b)$.

Let us briefly discuss our choices of renormalisation schemes.
It is convenient to perform the calculation in a first step in a
renormalisation scheme where the bottom quark is renormalised in the pole scheme (since we have $q^2=m_b^2$ for the external momentum) and charm quark mass in the $\overline{\rm MS}$ scheme.
In a next step we transform the bottom quark mass in the quantity $z=m_c^2/m_b^2$ to the $\overline{\rm MS}$ scheme and introduce separate renormalisation scales $\mu_c$ and $\mu_b$ for the masses and $\mu_1$ for $\alpha_s$. For
our numerical analysis we identify these scales (i.e. we set
$\mu_1=\mu_c=\mu_b$) and adapt $\mu_1=4.2$~GeV as central value.  We estimate
the uncertainty due to unknown higher order corrections by varying $\mu_1$
between $2.1$~GeV and $8.4$~GeV. Next
we introduce three choices for the overall factor $m_b^2$ (see Eqs.~\eqref{eq:tu-td} and \eqref{eq::diff}): We
either keep it in the pole scheme or we transform it to the
$\overline{\rm MS}$ or potential-subtracted (PS) scheme~\cite{Beneke:1998rk}.  In the
following we present numerical results in these renormalisation
schemes.

In Eq.~(\ref{eq::diff}) we insert NNLO corrections to the leading-CKM matching
coefficients {$\vec{F}^{cu}$} and {$\vec{F}^{cd}$} while only NLO corrections are used for the CKM-suppressed matching coefficient {$\vec{F}^{cc}$}, {$\vec{F}^{uu}$}, {$\vec{\tilde F}^{cu}$}, {$\vec{F}^{cs}$} and {$\vec{F}^{ud}$}.  Note that the latter only
contributes to about $0.005\%$ and $0.06\%$
in the $\overline{\rm MS}$ and PS scheme, respectively.
Also the penguin contribution, which we include to LO
are well below the per mille level.
Since we do not include {$\vec{F}^{cc}$}, {$\vec{F}^{uu}$}, {$\vec{\tilde F}^{cu}$}, {$\vec{F}^{cs}$} and {$\vec{F}^{ud}$} to NNLO we furthermore use the approximation $|V_{ud}|=1$, {$|V_{ub}|=0$} and $|V_{cd}|=0$ at NNLO.

In our analysis we do not include contributions from dimension-seven
operators. The LO coefficients are available from 
Refs.~\cite{King:2021xqp,Lenz:2022rbq}, but the matrix elements are completely unknown.
Confronting our results with experimental data probes the size of these terms.

In the following we present our predictions for $\tau(B^+)/\tau(B_d^0)$
including the uncertainties from the variation of $\mu_1$ (``scale''), from
the bag parameters (``bag''), from $|V_{cb}|$ and the remaining input
parameters as given in Tab.~\ref{tab::input}.  For the three renormalisation
schemes introduced above we obtain
\begin{align}
\frac{\tau(B^+)}{\tau(B_d^0)} &= {1.073^{+0.003}_{\scalebox{2}[1]{-}  0.024}}_{\text{scale}} \pm 0.018_{\text{bag}} \pm 0.002_{|V_{cb}|} \pm 0.002_{\text{input}}\, (\overline{\text{MS}})\,, \nonumber\\[5pt]
\frac{\tau(B^+)}{\tau(B_d^0)} &= {1.071^{+0.008}_{\scalebox{2}[1]{-}  0.020}}_{\text{scale}} \pm 0.018_{\text{bag}} \pm 0.002_{|V_{cb}|} \pm 0.001_{\text{input}}\, (\text{PS})\,, \nonumber\\[5pt]
\frac{\tau(B^+)}{\tau(B_d^0)} &= {1.066^{+0.013}_{\scalebox{2}[1]{-} 0.02{7}}}_{\text{scale}}  \pm 0.017_{\text{bag}} \pm 0.002_{|V_{cb}|} \pm 0.001_{\text{input}}\, (\text{pole})\,.
\label{eq::tau_ra}
\end{align}
We observe that the central values from the $\overline{\rm MS}$ and PS schemes
are very close with similar uncertainty estimates. The prediction from the pole
scheme is slightly smaller and shows a larger uncertainty.
The dominant contributions to the uncertainties come from the scale variation and
the bag parameters; the parametric uncertainties are much smaller.

\begin{figure}[t]
    \begin{center}
      \includegraphics[width=0.9\linewidth]{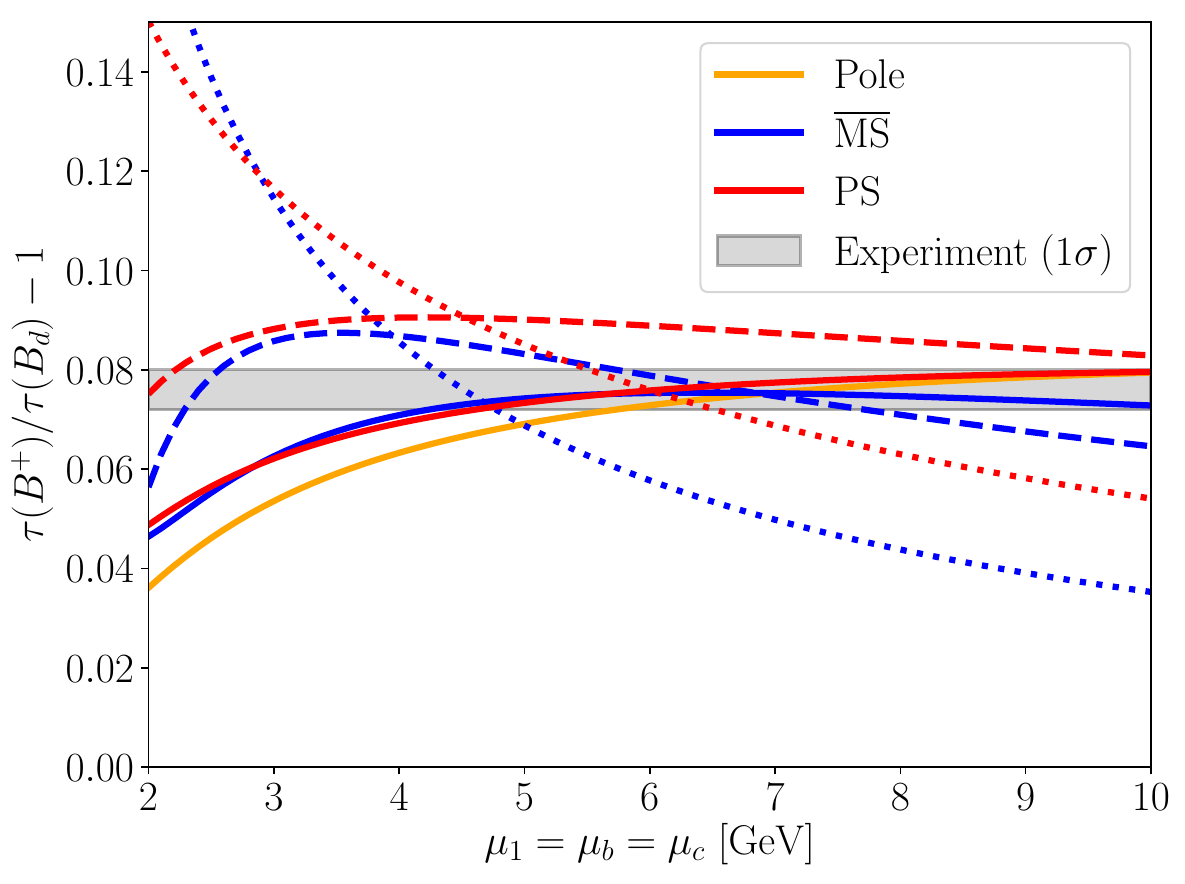}
    \end{center}
\caption{\label{fig::tau_ra}$\tau(B^+)/\tau(B_d^0)-1$ as a function of $\mu_1$ for different renormalisation schemes, see legend. The current experimental results with a 1$\sigma$ uncertainty are shown as gray band.}
\end{figure}

In Fig.~\ref{fig::tau_ra} we show $\tau(B^+)/\tau(B_d^0)-1$ as a function of
$\mu_1$ between $2$~GeV and $10$~GeV.  For the $\overline{\rm MS}$ and PS
schemes we show LO, NLO and NNLO results as dotted, dashed and solid lines.
For comparison also the NNLO predictions in the pole scheme are shown. The
current experimental result is shown as gray band.  We observe that the LO
result has a strong dependence on $\mu_1$ which is significantly reduced when
going to NLO. In the range of $\mu_1\in[2.1~\mbox{GeV},8.4~\mbox{GeV}]$ the
step from NLO to NNLO does not lead to an improvement: There is a reduction
from 20\% to 19\% in the $\overline{\rm MS}$ scheme and even an increase from
$7\%$ to $20\%$ in the PS scheme.  However, the solid curves are much closer
together than the dashed curves which indicates a significant reduction of the
renormalisation scheme dependence at NNLO. Also the agreement with the
experimental results is improved at NNLO.

For our final prediction we only consider the $\overline{\rm MS}$ and PS
schemes. In each scheme we add in quadrature upper and lower uncertainties,
symmetries the result and perform an average across the two schemes.
This leads to
\begin{equation}
  \frac{\tau(B^+)}{\tau(B_d^0)} = 1.072 \pm 0.024\,,
  \label{eq:tau_meson_final_result}
\end{equation}
which has to be compared to the experimental value~\cite{HFLAV:2024ctg} 
\begin{equation}
  \left(\frac{\tau(B^+)}{\tau(B_d^0)}\right)^{\text{exp}} = 1.076 \pm 0.004\,.
\end{equation}
    The experimental uncertainty is  smaller than the one
from the theory prediction by a factor of six.  We observe good agreement, well within the
cited uncertainties.  Since it is expected that beyond-the-Standard
Model effects the the lifetime ratio are small, 
we interpret the good agreement as evidence for the validity of the
heavy-quark expansion.

Further reduction of the theory uncertainty can be achieved by improving the
predictions of the bag parameters and by computing further higher-order
corrections in perturbation theory.

It is interesting to compare our NNLO prediction from
Eq.~(\ref{eq:tau_meson_final_result})
to previous predictions which were based on NLO calculations.
Ref.~\cite{Lenz:2022rbq} cites the result
\begin{equation}
  \frac{\tau(B^+)}{\tau(B_d^0)} = \SI{1.086(22)}{}\,.
\end{equation}
which is based on an older determination of the non-perturbative matrix elements.
It also includes the power-suppressed contributions from dimension-seven
operators in the vacuum insertion approximation, which sets the colour-octet matrix elements to zero.
Since the colour-singlet operators have small coefficients, similar to the dimension-six case (see \eqsto{eq:tau_individual_contributions_MS}{eq:ind_pole} below), this results in a small contribution.
We find agreement within the respective uncertainties. The
uncertainty given in Ref.~\cite{Lenz:2022rbq} is smaller than the one given in
Eq.~(\ref{eq:tau_meson_final_result}) since the scale variation was done in a
different way and over a smaller interval.

For convenience we present in the following numerical results for
$\tau(B^+)/\tau(B_d^0)$ without inserting numerical results for the
bag parameters. This enables to obtained improved predictions once
more precise results for the bag parameters become available.
We parametrise $\tau(B^+)/\tau(B_d^0)$ as
\begin{eqnarray}\label{eq:tau_individual_contributions}
  \frac{\tau(B^+)}{\tau(B_0)} 
  &=& 1 
      + F_{B_1} B_1
      + F_{B_2} B_2
      + F_{\epsilon_1} \epsilon_1
      + F_{\epsilon_2} \epsilon_2
\end{eqnarray}
and provide results for the coefficients $F_i$
using the same prescription as for Eq.~(\ref{eq::tau_ra}).
For the three renormalisation schemes we obtain:
\begin{alignat}{4}
F_{B_1}^{\overline{\text{MS}}} &= {\phantom{-}0.027^{+0.003}_{\scalebox{2}[1]{-}  0.015}}_{\text{scale}} &&\pm 0.00{1}_{|V_{cb}|} &&\pm 0.001_{\text{input}}\,,\nonumber\\
F_{B_2}^{\overline{\text{MS}}} &= {-0.00{6}^{+0.00{3}}_{\scalebox{2}[1]{-}  0.008}}_{\text{scale}} &&\pm 0.000_{|V_{cb}|} &&\pm 0.000_{\text{input}}\,,\nonumber\\
F_{\epsilon_1}^{\overline{\text{MS}}} &= {-0.56{8}^{+0.028}_{\scalebox{2}[1]{-}  0.001}}_{\text{scale}} &&\pm 0.01{5}_{|V_{cb}|} &&\pm 0.010_{\text{input}}\,,\nonumber\\
F_{\epsilon_2}^{\overline{\text{MS}}} &= {\phantom{-}0.11{6}^{+0.002}_{\scalebox{2}[1]{-}  0.003}}_{\text{scale}} &&\pm 0.00{3}_{|V_{cb}|} &&\pm 0.002_{\text{input}}\,\label{eq:tau_individual_contributions_MS},
\end{alignat}
\begin{alignat}{4}
F_{B_1}^{\text{PS}} &= {\phantom{-}0.02{5}^{+0.007}_{\scalebox{2}[1]{-}  0.013}}_{\text{scale}} &&\pm 0.00{1}_{|V_{cb}|} &&\pm 0.001_{\text{input}}\,,\nonumber\\
F_{B_2}^{\text{PS}} &= {-0.00{6}^{+0.002}_{\scalebox{2}[1]{-}  0.006}}_{\text{scale}} &&\pm 0.000_{|V_{cb}|} &&\pm 0.000_{\text{input}}\,,\nonumber\\
F_{\epsilon_1}^{\text{PS}} &= {-0.57{5}^{+0.014}_{\scalebox{2}[1]{-}  0.001}}_{\text{scale}} &&\pm 0.01{5}_{|V_{cb}|} &&\pm 0.011_{\text{input}}\,,\nonumber\\
F_{\epsilon_2}^{\text{PS}} &= {\phantom{-}0.11{6}^{+0.005}_{\scalebox{2}[1]{-}  0.003}}_{\text{scale}} &&\pm 0.00{3}_{|V_{cb}|} &&\pm 0.002_{\text{input}}\,. \label{eq:ind_ps}
\end{alignat}
\begin{alignat}{4}
F_{B_1}^{\text{pole}} &= {\phantom{-}0.0{22}^{+0.010}_{\scalebox{2}[1]{-}  0.017}}_{\text{scale}} &&\pm 0.000_{|V_{cb}|} &&\pm 0.001_{\text{input}}\,,\nonumber\\
F_{B_2}^{\text{pole}} &= {-0.00{7}^{+0.002}_{\scalebox{2}[1]{-}  0.006}}_{\text{scale}} &&\pm 0.000_{|V_{cb}|} &&\pm 0.000_{\text{input}}\,,\nonumber\\
F_{\epsilon_1}^{\text{pole}} &= {-0.55{4}^{+0.035}_{\scalebox{2}[1]{-}  0.012}}_{\text{scale}} &&\pm 0.01{4}_{|V_{cb}|} &&\pm 0.010_{\text{input}}\,,\nonumber\\
F_{\epsilon_2}^{\text{pole}} &= {\phantom{-}0.10{9}^{+0.008}_{\scalebox{2}[1]{-}  0.006}}_{\text{scale}} &&\pm 0.00{3}_{|V_{cb}|} &&\pm 0.002_{\text{input}}\,. \label{eq:ind_pole}
\end{alignat}

Comparing the results across the three schemes, the pole scheme has the largest perturbative uncertainty for all coefficients, which is consistent with the expected poor convergence of the perturbative series in this scheme \cite{Beneke:1994sw,Bigi:1994em}. For each scheme, the four coefficients differ in the relative size of the perturbative uncertainty. Interestingly, the coefficients of the larger bag parameters $B_{1}$ and $B_2$ also have a larger relative perturbative uncertainty, with that of $B_2$ exceeding 100\%.
The coefficient $F_{B_2}$ is small compared to $F_{B_1}$ because the contribution to $F_S^{cd}$ vanishes to
LO whereas it gives a sizeable contribution to $F^{cd}$. 
Therefore, more accurate results of the lifetime ratio could be achieved by calculating higher orders in perturbation theory of selected matching coefficients.



\section{\label{sec::phen_charm}Phenomenology for $\mathbf{D}$ mesons}

For the numerical evaluation of lifetime ratios in the $D$ meson
system we adapt our results obtained for $B$ meson decays.  In a
first step we set all contributions where charm quarks appear as
virtual particles in the gluon propagator to zero.  Next we replace
$m_b$ by $m_c$ and interpret the quantity $z=m_c^2/m_b^2$ as
$z_s=m_s^2/m_c^2 $ and adapt the CKM factors such that we end up with
matching coefficients expressed in terms of quantities in
$n_f=4$-flavour QCD, in particular $\alpha_s^{(4)}(\mu_1)$.  It is
convenient to evaluate the matching coefficients $C_1, C_2, \ldots$ in
the 5-flavour theory. For this reason one should use the relation between
$\alpha_s^{(4)}$ and $\alpha_s^{(5)}$ before the numerical evaluation
of the lifetime ratio.

However, since a precision analysis of $D$ meson lifetimes in beyond the scope of this work
we  neglect all threshold effects
between $n_f=4$- and $n_f=5$-flavour QCD, i.e.,
we use $m_c^{(5)}$ and $\alpha_s^{(5)}$ in the
routines which evaluate the lifetime differences.
The numerical effect of this approximation is expected to be very small.
Furthermore we must set $z_s=0$, because effects of $m_s \neq 0$ correspond to 
higher-dimensions in the HQE.

\begin{figure}[t]
    \centering
    \begin{tabular}{cc}
         \includegraphics[width=0.45\linewidth]{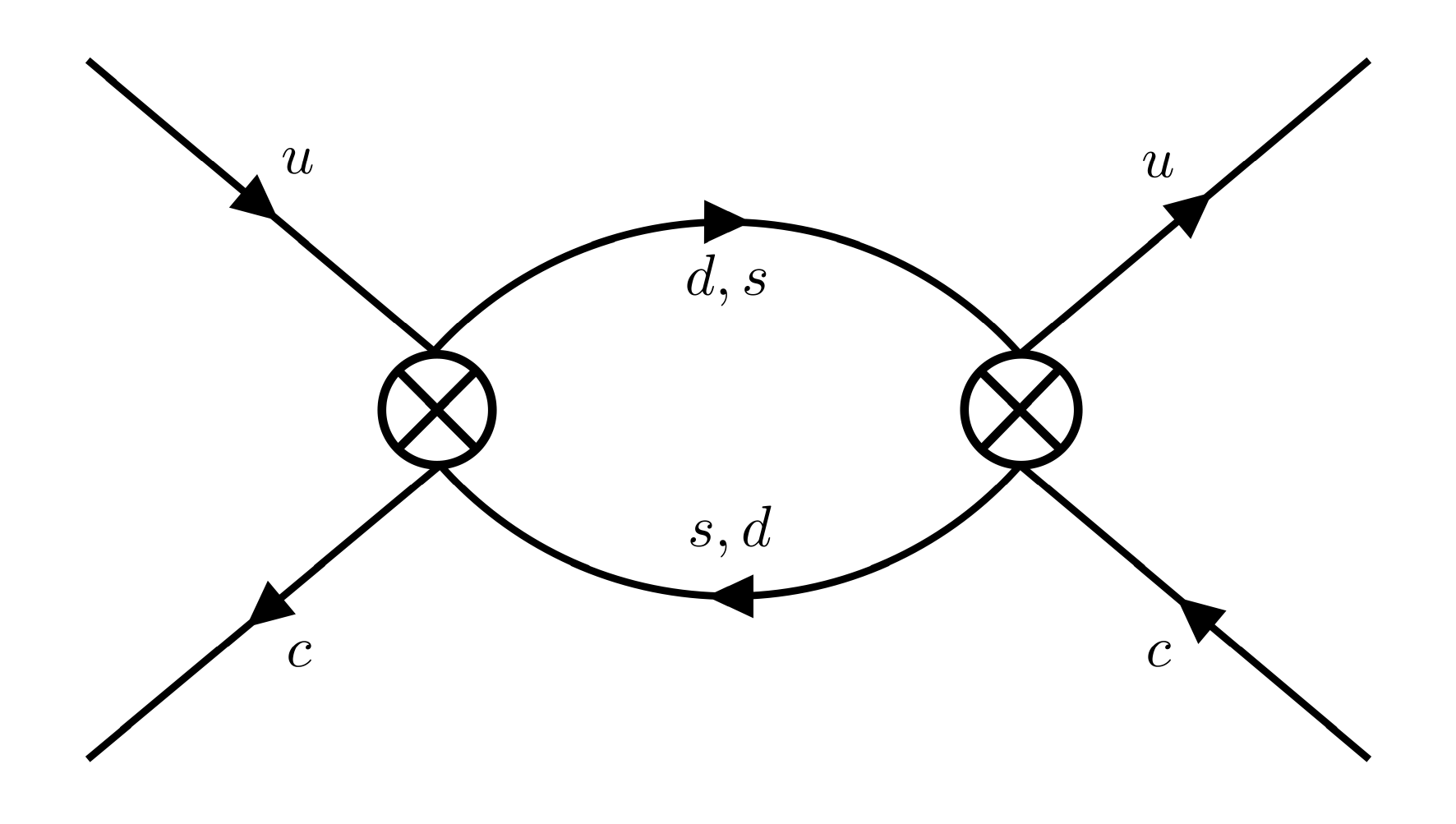}
         &\includegraphics[width=0.45\linewidth]{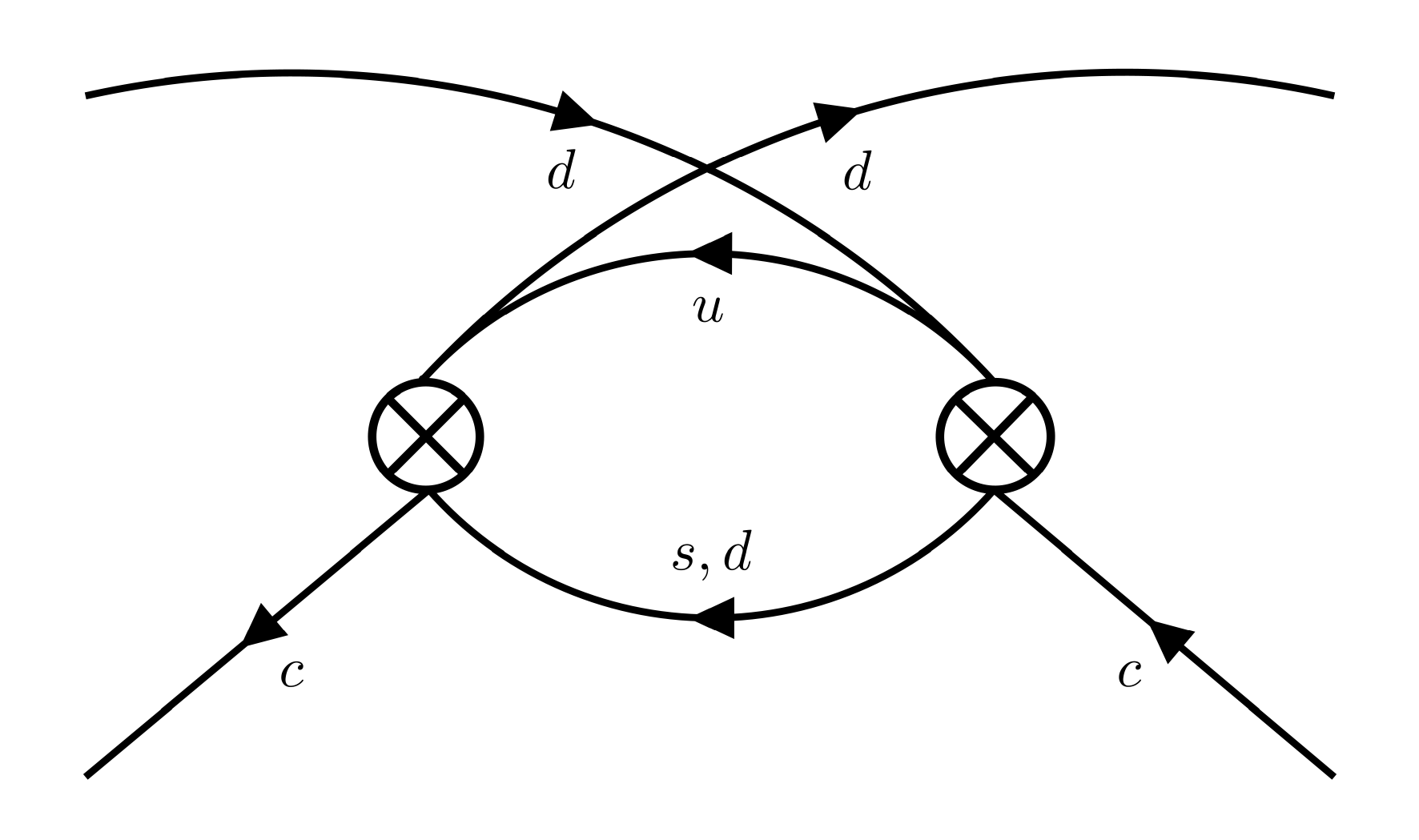}\\
         \hspace{4cm}\includegraphics[width=0.5\linewidth]{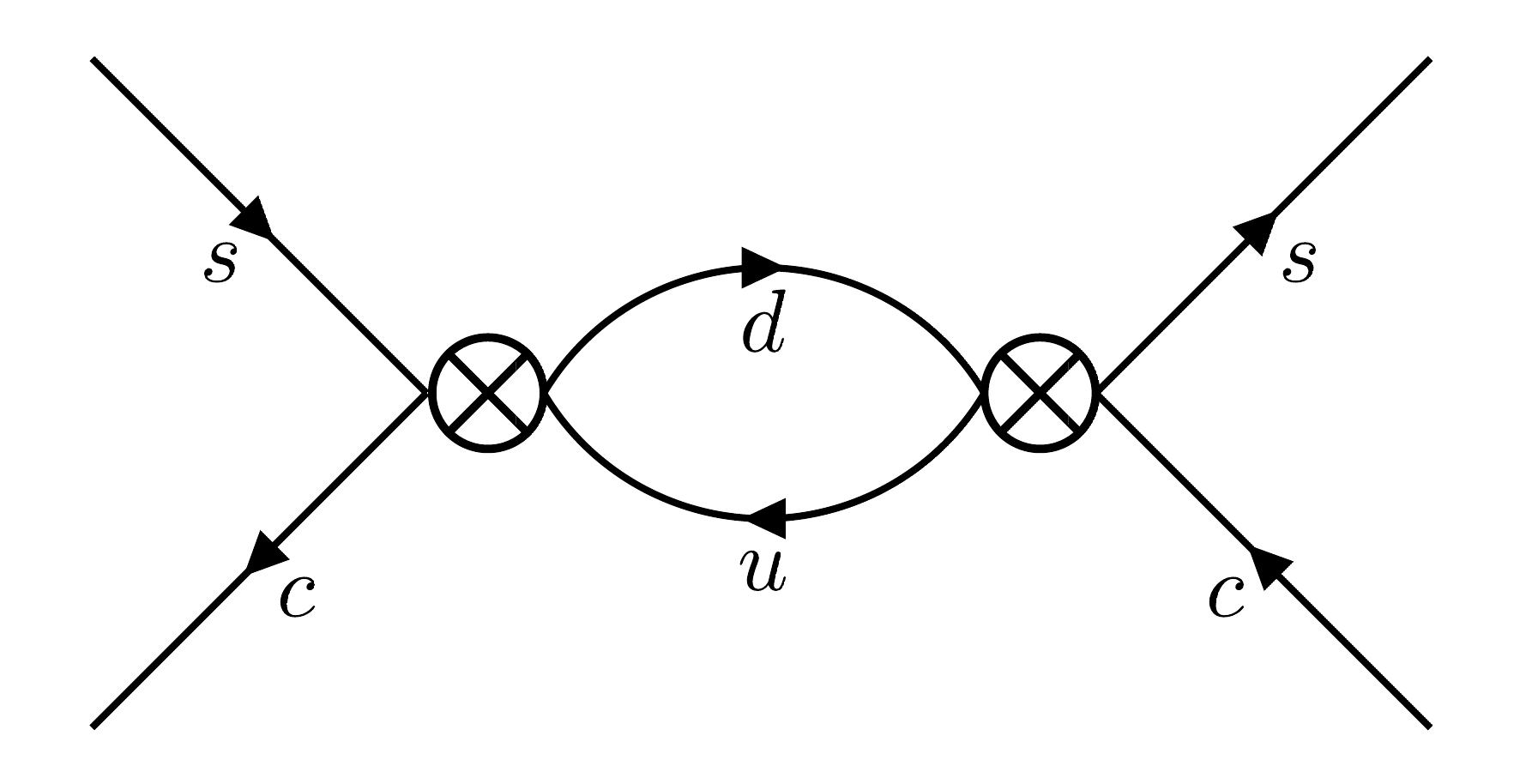}\hspace{-4cm}
    \end{tabular}
    \caption{
    Weak annihilation contribution to $\Gamma(D^0)$ (top left), Pauli interference contribution to $\Gamma(D^+)$ (top right), and weak annihilation contribution to $\Gamma(D_s^+)$ (bottom) in the leading order of QCD. The crosses represent the insertion of $|\Delta C| = 1$ operators from the effective Hamiltonian of weak charm decays. CKM-suppressed contributions are not shown.} 
    \label{fig:wa-pi-LO-D-meson}
\end{figure}

We consider the isospin doublet $(D^+,D^0)$ and the V-spin doublet $(D_s^+,D^0)$, and compute the ratios $\tau(D^+)/\tau(D^0)$ and ${\tau(D_s^+)/\tau(D^0)}$. We stress that the CKM-suppressed and CKM-favoured contributions to $\Gamma(D_s^+)$ are interchanged with respect to $\Gamma(D^+)$ and $\Gamma(B^+)$. The diagrams contributing at LO in QCD to the decay rates of $D$ mesons are displayed in Fig.~\ref{fig:wa-pi-LO-D-meson}.

\subsection{\label{subs:Dplus/D0} The lifetimes ratio ${\tau(D^+)/\tau(D^0)}$}

In the case of charmed-hadron lifetime ratios 
the contributions of the dimension-7 operators in the HQE of the decay width, as in Eq.~\eqref{eq:hqe}, are expected to have a larger impact on the theoretical predictions~\cite{Lenz:2013aua} compared to the bottom case, because $\Lambda_{\rm QCD}/m_c\sim 0.3$ is larger than $\Lambda_{\rm QCD}/m_b\sim 0.1$. As mentioned above, we deviate from Ref.~\cite{Lenz:2013aua} by neglecting these contributions and using the
 experimental measurement as an estimate of the size of the $1/m_c^4$ terms. Placing such bounds helps to assess the size of the corresponding terms in the $b$ system, whose impact should be three times smaller there.

\begin{table}[t]
  \begin{center}
  {\scalefont{0.9}
    \renewcommand{\arraystretch}{1.2}
    \begin{tabular}{rcll | rcll}
      \hline
       $f_{D}$ &=& $\SI{0.2120(7)}{GeV}$ & \cite{FLAG:2024oxs}
      &
      $M_{D}$ &=& $\SI{1869.66(05)}{MeV}$ & \cite{ParticleDataGroup:2024cfk} 
      \\
      $\tau(D^0)$ &=& $\SI{.4103(10)}{\pico\second}$ & \cite{ParticleDataGroup:2024cfk}
      &
      $\tau(D^+)$ &=& $\SI{1.033(5)}{\pico\second}$ & \cite{ParticleDataGroup:2024cfk}
      \\
      $\tau(D^+)/\tau(D^0)$ &=& $2.510(13)(07)$ & \cite{Belle-II:2021cxx}
      &$\tau(D_s^+)$ &=& $\SI{.5012(22)}{\pico\second}$ & \cite{ParticleDataGroup:2024cfk}\\
      $\tau(D_s^+)/\tau(D^0)$ &=& $1.222(6)$ & \cite{ParticleDataGroup:2024cfk}
      &${B(D_s^+\to \tau^+ \nu_\tau)}$ &=& {$5.39(9)\%$} & \cite{ParticleDataGroup:2024cfk}
      \\
      {$B(D^0\to Xe^+ \nu_e)$} &=& {$6.49(11)\%$} & \cite{ParticleDataGroup:2024cfk}
      &{$B(D_s^+\to X e^+ \nu_e)$} &=& {$6.33(15)\%$} & \cite{ParticleDataGroup:2024cfk}
      \\
      $f_{D_s}$ &=& $\SI{0.2499(5)}{GeV}$ & \cite{FLAG:2024oxs}
      &$f_{D_s}/f_{D}$ &=& $1.1783(16)$ & \cite{FLAG:2024oxs}
      \\
      $|V_{us}|$ &=& $0.22504^{+0.00020}_{\scalebox{2}[1]{-}  0.00022}$ &
                                                                             \cite{CKMfitter}
      &
      
      $|V_{cs}|$ &=& $0.973497^{+0.000057}_{\scalebox{2}[1]{-}  0.000054}$ &
                                                                             \cite{CKMfitter}
      \\
      \hline
    \end{tabular}
    }
  \end{center}
  \caption{\label{tab::input-D-mesons}Input parameters for the numerical analysis. 
 }
\end{table}

Following the procedure that leads to Eq.~\eqref{eq::diff}, we obtain \cite{Lenz:2013aua}
\begin{eqnarray}
    & \Gamma(D^0) - \Gamma(D^+) = \frac{G_F^2 m_c^2}{12\pi}f_D^2 M_D \left(\vec{F}^{s^\prime d^\prime}_{{D}}-|V_{ud}|^2 \vec{F}_{{D}}^{s^\prime u}\right)\cdot \vec{B},
    \label{eq::diff-D-meson}
\end{eqnarray}
where $F^{qd^\prime}_{{D}}=|V_{ud}|^2 F^{qd}_{{D}}+|V_{us}|^2 F^{qs}_{{D}}$ and $F^{s^\prime q}_{{D}}=|V_{cd}|^2 F^{d q}_{{D}}+|V_{cs}|^2 F^{sq}_{{D}}$. Here, the indices $qq^\prime$ in the notation of the Wilson coefficients refer to the intermediate $qq^\prime$ quark pair. When we adapt our results for the $B$ mesons to the $D$ system, we set $x=0$ and identify {$F^{cu}=F^{sd}_{{D}}=F^{ds}_{{D}}=F^{dd}_{{D}}=F^{ss}_{{D}}$} and $F^{cd}=F^{su}_{{D}}=F^{du}_{{D}}$.

For the numerical evaluation we use the input parameters given in Tab.~\ref{tab::input-D-mesons} and the bag
parameters of the $D$ mesons obtained from QCD sum rules in
Ref.~\cite{Black:2024bus},
\begin{align}
  B_1^{{D}} &= 0.875^{+0.07}_{\scalebox{2}[1]{-} 0.044}\,,\nonumber\\
  B_2^{{D}} &= 0.862^{+0.138}_{- 0.078}\,, \nonumber\\
  \epsilon_1^{{D}} &= -0.122^{+0.033}_{- 0.042}\,,\nonumber\\
  \epsilon_2^{{D}} &= 0.0002^{+0.0148}_{- 0.0197}\,.
\label{eq:bag_decay_input_D_meson}
\end{align}
They are defined at the scale $\mu_0=\SI{3}{GeV}$.

For our final prediction we consider only the $\overline{\text{MS}}$ scheme. We adapt $\mu_1=\SI{3}{GeV}$ as our central value, and let $\mu_1$ vary between $\SI{1.5}{GeV}$ and $\SI{6}{GeV}$ to estimate the uncertainty due to higher order corrections. 
Explicitly, we find
\begin{eqnarray}
  \left(\frac{\tau(D^+)}{\tau(D^0)}\right)^{\rm LO} &=& {2.073^{+1.179}_{\scalebox{2}[1]{-}  0.489}}_{\text{scale}}\,,\nonumber\\
  \left(\frac{\tau(D^+)}{\tau(D^0)}\right)^{\rm NLO} &=& {1.927^{+0.002}_{\scalebox{2}[1]{-}  0.702}}_{\text{scale}}\,,\nonumber\\
  \left(\frac{\tau(D^+)}{\tau(D^0)}\right)^{\rm NNLO} &=& {1.737^{+0.082}_{\scalebox{2}[1]{-}  0.182}}_{\text{scale}} \pm 0.185_{\text{bag}} \pm 0.014_{\text{input}}\,.
  \label{eq:tau_D_meson_final_result}
\end{eqnarray}
Here and in the following we omit the uncertainties from bag parameters and input in the LO and NLO numbers, because they 
do not change with the order of the perturbative series.

Recently the bag parameters for $\tau(D_s^+)/\tau(D^0)$ have 
been computed with lattice QCD. To apply these to ${\tau(D^+)}/\tau(D^0)$ one 
must trust in U-spin symmetry; which in all lattice calculations for $|\Delta B|=2$ bag parameters 
is found to hold to an excellent level. The dominant effect of U-spin breaking in hadronic matrix elements is captured by the ratio of the decay constants, so that we assume that $D_s^+$ matrix elements differ from their $D^+$ counterparts by a factor of $f_{D_s}^2/f_D^2\sim 1.4$. In $B$ physics the large U-spin breaking in decay constants has been ascribed to large chiral logarithms which are absent in the bag parameters~\cite{Kronfeld:2002ab}. Thus we use the same lattice bag parameters for both lifetime ratios~\cite{Black:2026rbz,Black:2026dzp}:
\begin{eqnarray}
  B_1^{{D}} &= 1.103\pm 0.010\,,\nonumber\\
  B_2^{{D}} &= 0.9754\pm 0.0071\,, \nonumber\\
  \epsilon_1^{{D}} &= -0.2427 \pm 0.0071\,,\nonumber\\
  \epsilon_2^{{D}} &= -0.0289\pm 0.0008\,,
\label{eq:bag_decay_input_D_meson_lat}
\end{eqnarray}
and obtain
\begin{eqnarray}
  \left(\frac{\tau(D^+)}{\tau(D^0)}\right)^{\rm LO} &=& {2.787^{+1.803}_{\scalebox{2}[1]{-}  0.757}}_{\text{scale}}\, (\overline{\text{MS}})\,,\nonumber\\
  \left(\frac{\tau(D^+)}{\tau(D^0)}\right)^{\rm NLO} &=& {2.590^{+0.001}_{\scalebox{2}[1]{-}  0.988}}_{\text{scale}}\, (\overline{\text{MS}})\,,\nonumber\\
  \left(\frac{\tau(D^+)}{\tau(D^0)}\right)^{\rm NNLO} &=& {2.344^{+0.094}_{\scalebox{2}[1]{-}  0.236}}_{\text{scale}} \pm 0.035_{\text{bag}} \pm 0.026_{\text{input}}\, (\overline{\text{MS}})\,.
  \label{eq:tau_D_meson_final_result_lat}
\end{eqnarray}
Symmetrising the scale uncertainty in \eq{eq:tau_D_meson_final_result_lat} and adding the three uncertainties in quadrature 
gives our final prediction 
\begin{eqnarray}
 \left(\frac{\tau(D^+)}{\tau(D^0)}\right)^{\rm NNLO} &=& 2.344 \pm 0.170 \,.\label{eq:dfinal}
\end{eqnarray}

This has to be compared to the experimental value~\cite{Belle-II:2021cxx} 
\begin{equation}
  \left(\frac{\tau(D^+)}{\tau(D^0)}\right)^{\text{exp}} = 2.510 \pm 0.015\,.
  \label{eq:tau_D_meson_exp}
\end{equation}
When we use the determination of the bag parameters from Ref.~\cite{Black:2024bus}, the comparison between the central values of our theoretical prediction of the lifetimes ratio and the quoted experimental value implies a magnitude of slightly less than 50\% for the $1/m_c^4$ suppressed terms, that is $\tilde \Gamma_7$ in Eq.~\eqref{eq:hqe}. This is in agreement with the estimate in Ref.~\cite{Lenz:2013aua}. The situation, however, changes considerably when we implement in our analysis the results presented recently in Refs.~\cite{Black:2026dzp,Black:2026rbz}. Here, a comparison between the central value in Eq.~\eqref{eq:dfinal} and the experimental determination in Eq.~\eqref{eq:tau_D_meson_exp} implies a contribution from the dimension-7 operators of slightly less than 10\%. This is due to the fact that the bag parameters differ significantly between the two determinations that we implement. In particular, as pointed out in Ref.~\cite{Black:2026dzp}, the difference in the value of $\epsilon_1^{\overline{\text{MS}}}$ is the more phenomenologically relevant, given that the Wilson coefficient associated to this matrix element is the largest~\cite{King:2021xqp}, as can be seen from Eq.~\eqref{eq:tau_individual_contributions_MS}.

\subsection{\label{subs:Dsplus/D0} The lifetimes ratio $\tau(D_s^+)/\tau(D^0)$}

Unlike the decay of the $D^+$ meson, the leading-CKM contribution to $\Gamma(D_s^+)$ comes from WA diagrams, as shown in Fig.~\ref{fig:wa-pi-LO-D-meson}.

$(D_s^+,D^0)$ is a doublet of the V-spin symmetry which rotates $u$ and $s$ quarks into each other. 
Because of $m_s-m_u\sim 0.3\;\Lambda_{\rm QCD}$ it is broken by 30\% from soft QCD dynamics, so that
the V-spin singlet terms in $\mathcal{T}_{sing}$ do not cancel exactly in the lifetimes difference of the two mesons.
{Furthermore, V-spin breaking implies \mbox{$\langle D_s^+ | {\{Q,T\}^{s,u}} | D_s^+ \rangle \neq \langle D^0 | {\{Q,T\}^{u,s}} | D^0 \rangle$}, so that $\tau(D_s^+)/\tau(D^0)-1$ cannot be solely expressed in terms of matrix elements of differences like $Q^{s}-Q^{u}$, cf.\ \eq{eq:ome} for the isospin analogue. In view of the remark after \eq{eq:tau_D_meson_final_result} we incorporate the described V-spin breaking by using $f_{D_s}^2$ in the normalisation of the $\langle D_s^+| \ldots | D_s^+\rangle $ matrix elements while taking $f_D^2$ for 
$\langle D^0| \ldots | D^0\rangle $. 

Other corrections which must be taken into account are described and analysed in detail in Ref.~\cite{Lenz:2013aua}: There are CKM-unsuppressed contributions to the WA diagram in \fig{fig:wa-pi-LO-D-meson} from a lepton pair in the loop. $D_s^+ \to\tau^+\nu_\tau$ defies the HQE because of 
$m_{D_s} \sim m_{\tau}$ but this contribution can be taken care of by subtracting $\Gamma (D_s^+ \to\tau^+\nu_\tau)$ from $\Gamma_{\rm tot}$ using data on the branching fraction. We also include the semi-leptonic decays 
$D_s^+\to X \ell  \nu_\ell$ and 
$D^0 \to X \ell  \nu_\ell$ in this way and replace $\Gamma_{\rm tot}$ by the hadronic rate 
\begin{align}
    \Gamma_{\rm had} & \equiv \Gamma_{\rm tot} - \Gamma_{\mbox{\tiny(semi-)lept}} \no
\end{align}
amounting to
\begin{align}
    \frac{\tau(D_s^+)}{\tau(D^0)} -1 & =\, 
     \tau(D_s^+) \Bigg[ \Gamma_{\rm had} (D^0) - \Gamma_{\rm had} (D_s^+)  - \Gamma(D_s^+ \to\tau^+\nu_\tau)\, + \nonumber\\
      & \hspace{3cm}\sum_{\ell=e,\mu} \lt( \Gamma (D^0 \to X\ell \nu) - \Gamma (D_s^+ \to X \ell \nu) \rt) \Bigg] \nn
      &= \, \tau(D_s^+) \lt[ \Gamma_{\rm had} (D^0) - \Gamma_{\rm had} (D_s^+) \rt] - 
      B( D_s^+ \to\tau^+\nu_\tau) \nn 
      & \qquad\qquad +  \sum_{\ell=e,\mu} \lt( {\frac{\tau(D_s^+)}{\tau(D^0)}} 
             B(D^0 \to X\ell \nu) - B(D_s^+ \to X \ell \nu)
      \rt) \nn 
      &=\, \tau(D_s^+) \lt[ \Gamma_{\rm had} (D^0) - \Gamma_{\rm had} (D_s^+) \rt] 
           - {0.0219 \pm 0.0041} \label{eq:ghad}
\end{align}
where PDG values \cite{ParticleDataGroup:2024cfk} (assuming equal semi-leptonic branching ratios for $e$ and $\mu$) 
have been used in the last line. The first term in \eq{eq:ghad} defines $\lt. \tau(D_s^+)/\tau(D^0)\rt|_{\rm had}$ which 
we calculate in this paper.

Thus we need
\begin{eqnarray}
    & \Gamma_{{\rm had}}(D^0) - \Gamma_{{\rm had}}(D_s^+) = \frac{G_F^2 m_c^2}{12\pi}f_D^2 M_D \left[ \vec{F}_{{D}}^{s^\prime d^\prime}-{\frac{f_{D_s}^2}{f_D^2}} |V_{cs}|^2 \left( |V_{ud}|^2 \vec{F}_{{D}}^{du}
    {+|V_{us}|^2 \vec{F}_{{D}}^{{su}}} \right) \right]\cdot \vec{B}, \quad
    \label{eq::diff-Ds-meson}
\end{eqnarray}
where the coefficients $\vec{F}_{{D}}^{du}$ are extracted from the coefficients $\vec{F}^{cu}$ in an analogous way to the $\vec{F}^{cs}$ contributions in Eq.~\eqref{eq:tu-td} and $F^{cd}=F_D^{su}$}.
There are no  CKM-favoured PI contributions to $ \tau(D^+)/\tau(D^0)$; since WA diagrams are much smaller, this lifetime ratio is
close to 1. The last term in \eq{eq::diff-Ds-meson} includes the CKM-suppressed PI contribution to $\Gamma (D_s^+)$. We confirm the finding of Ref.~\cite{Lenz:2013aua} that this term is numerically smaller than the WA contribution but non-negligible.

For our final prediction we consider only the $\overline{\text{MS}}$ scheme. We adapt $\mu_1=\SI{3}{GeV}$ as our central value, and let $\mu_1$ vary between $\SI{1.5}{GeV}$ and $\SI{6}{GeV}$ to estimate the uncertainty due to higher order corrections. 
Furthermore, we set $m_s=0$ at LO, NLO and NNLO.
Since we assume exact $SU(3)$ symmetry, we take the values in Eq.~\eqref{eq:bag_decay_input_D_meson} for the bag parameters in Eq.~\eqref{eq::diff-Ds-meson}.
Explicitly, we find 
\begin{eqnarray}
  \left(\frac{\tau(D_s^+)}{\tau(D^0)}\right)^{\rm LO}_{\rm had} &=& {1.110^{+0.076}_{\scalebox{2}[1]{-}  0.035}}_{\text{scale}}\, (\overline{\text{MS}})\,,\nonumber\\
  \left(\frac{\tau(D_s^+)}{\tau(D^0)}\right)^{\rm NLO}_{\rm had} &=& {1.115^{+0.005}_{\scalebox{2}[1]{-}  0.091}}_{\text{scale}}\, (\overline{\text{MS}})\,,\nonumber\\
  \left(\frac{\tau(D_s^+)}{\tau(D^0)}\right)^{\rm NNLO}_{\rm had} &=& {1.157^{+0.074}_{\scalebox{2}[1]{-}  0.010}}_{\text{scale}} \pm 0.124_{\text{bag}} \pm 0.004_{\text{input}}\, (\overline{\text{MS}})\,.
  \label{eq:tau_D_s_meson_final_result}
\end{eqnarray}

When we use the results in Refs.~\cite{Black:2026dzp,Black:2026rbz} instead, we get 
\begin{eqnarray}
  \left(\frac{\tau(D_s^+)}{\tau(D^0)}\right)^{\rm LO}_{\rm had} &=& {1.288^{+0.176}_{\scalebox{2}[1]{-}  0.085}}_{\text{scale}}\, (\overline{\text{MS}})\,,\nonumber\\
  \left(\frac{\tau(D_s^+)}{\tau(D^0)}\right)^{\rm NLO}_{\rm had} &=& {1.275^{+0.003}_{\scalebox{2}[1]{-}  0.135}}_{\text{scale}}\, (\overline{\text{MS}})\,,\nonumber\\
  \left(\frac{\tau(D_s^+)}{\tau(D^0)}\right)^{\rm NNLO}_{\rm had} &=& {1.311^{+0.071}_{\scalebox{2}[1]{-}  0.008}}_{\text{scale}} \pm 0.013_{\text{bag}} \pm 0.006_{\text{input}}\, (\overline{\text{MS}})\,.
  \label{eq:tau_D_s_meson_final_result_lat}
\end{eqnarray}
In analogy to \eq{eq:dfinal} we find our final result for the choice of lattice bag parameters:
\begin{eqnarray}
  \left(\frac{\tau(D_s^+)}{\tau(D^0)}\right)^{\rm NNLO}_{\rm had} &=& 1.311 \pm 0.042
  \label{eq:dsfinal_1}
\end{eqnarray}

This has to be confronted with the corresponding quantity derived from the 
experimental value~\cite{ParticleDataGroup:2024cfk} 
\begin{equation}
  \left(\frac{\tau(D_s^+)}{\tau(D^0)}\right)^{\text{exp}} = 1.222 \pm 0.006\,. 
  \label{eq:tauDsovertauD0_total_exp}
\end{equation}

Subtracting the contributions from the (semi-)leptonic decays as described around \eq{eq:ghad}
one finds the experimental result for the hadronic contribution to the lifetime ratio as
\begin{equation}
    \left(\frac{\tau(D_s^+)}{\tau(D^0)}\right)^{\text{exp}}_{\text{had}} = {1.244} \pm 0.007\,. \label{eq:exphad}
\end{equation}

Similarly to what happens in the case of the ratio $\tau(D^+)/\tau(D_0)$, the results obtained with lattice QCD show a better agreement with the experimental value with respect to the results derived using HQET sum rules.
Contrary to the case of $\tau(D^+)/\tau(D_0)$, the difference between \eq{eq:exphad} and 
\eqsand{eq:tau_D_s_meson_final_result}{eq:tau_D_s_meson_final_result_lat}, is not only a measure of the suppressed terms of 
${\cal O} (1/m_c^4)$ and higher in the HQE and unknown N$^3$LO corrections, but also of the 
V-spin breaking terms beyond the factorisable contributions accounted for by the factor of $f_{D_s}^2/f_D^2$.
Comparing the central values, the experimental value in \eq{eq:exphad} differs from the NNLO 
results in \eqsand{eq:tau_D_s_meson_final_result}{eq:tau_D_s_meson_final_result_lat} by 
$0.087$ and $-0.067$, respectively, 
slightly favouring the lattice computation. Yet both values are smaller in magnitude than the experimental value of 0.244 of the calculated 
valence-quark effects, indicating that the HQE works at the expected level of accuracy. 

For completeness we finally add the (semi-)leptonic decay rates of \eq{eq:ghad} to \eq{eq:dsfinal_1} to find 
\begin{eqnarray}
  \left(\frac{\tau(D_s^+)}{\tau(D^0)}\right)^{\rm NNLO} &=& 1.289 \pm 0.042 .
  \label{eq:dsfinal}
\end{eqnarray}



\section{\label{sec::concl}Conclusions}

In this paper we have presented a new SM prediction for the lifetime ratio $\tau(B^+)/\tau(B_d^0)$ based on a 
calculation of next-to-next-leading order (NNLO) QCD corrections to the Wilson coefficients of the dimension-6 operators
appearing in the heavy quark expansion (HQE) of this ratio. This is the first such calculation for $\Delta B=0$ coefficients 
and constitutes an important step towards theory predictions of lifetime difference among $b$-flavoured hadrons which match the 
accuracy of the corresponding measurements. The computation has involved two-scale three-loop integrals which we have 
determined as deep expansions in the quark mass ratio $m_c/m_b$. In addition, we have calculated Cabibbo-suppressed contributions 
to the lifetime ratio at NLO.

We find sizable NNLO corrections, comparable to the NLO contribution, with a significant reduction of the dependence on the renormalisation scheme. The residual renormalisation scale dependence is satisfactory and, most importantly, the NNLO corrections 
combined with hadronic matrix elements from QCD sum rules \cite{Kirk:2017juj,Black:2024bus} bring the calculated $\tau(B^+)/\tau(B_d^0)$ into excellent agreement
with the experimental value. This suggest that contributions from yet poorly known dimension-7 terms of the HQE are small, 
so that the HQE indeed constitutes a valid and useful conceptual concept. We have presented our results in a form which permits the implementation of future, more precise computations of  the hadronic matrix elements in an easy and  straightforward way. 

Applying our results to the charm system, we present predictions for the lifetime splittings in the isospin doublet $(D^+,D^0)$ and 
the V-spin doublet $(D_s^+,D^0)$. The larger values of $\alpha_s(m_c)$ and $\lqcd/m_c$ compared to their counterparts in the $b$ system lead to larger uncertainties from neglected higher-order terms in the HQE. We combine our NNLO coefficients with both the matrix elements calculated with QCD sum rules \cite{Black:2024bus} and the ones computed on the lattice \cite{Black:2026dzp,Black:2026rbz}. In both cases our predictions agree with the experimental results within uncertainties. The difference between theoretical and experimental numbers can be viewed as an estimate of the $(\lqcd/m_c)^4$
terms of the HQE and in the case of $\tau (D_s^+)/\tau(D^0)$ also of unaccounted V-spin breaking terms. For the lattice-based 
prediction the difference amounts to $-11\%$ and ${27\%}$  of the calculated valence-quark contributions $\tau (D^+)/\tau(D^0)-1$ and
$\tau (D_s^+)/\tau(D^0)-1$, respectively,  
if one simply compares the central values of prediction and measurement. This is well below the expected  uncertainty of order $\lqcd/m_c \sim 0.3$ from the next term in the HQE. If one uses the sum-rule results for the matrix elements, this exercise gives larger differences of $-51\%$ and ${-36\%}$ for the two quantities.   

Our NNLO result supports the HQE as the correct method to calculate lifetime differences between heavy hadrons, for both the beauty and charm systems. Extending predictions to even higher orders of $\alpha_s(m_{b,c})$ and $\lqcd/m_{b,c}$ in combination with better 
hadronic matrix elements will reduce the theoretical uncertainties further with the aim to match the small experimental errors. 
In observables with essentially no sensitivity to physics beyond the SM, like the lifetime ratios studied in this paper, 
these calculation will permit to assess the final accuracy achievable with the HQE. With justified confidence in the method one can then 
apply the HQE to observables with high BSM sensitivity, especially those related to the decay width matrices of the $B_{d,s}$--$\bar B_{d,s}$ mixing complexes \cite{Nierste:2025muk}.   



\section*{Acknowledgements}  

This research was supported by the Deutsche Forschungsgemeinschaft (DFG,
German Research Foundation) under grant 396021762 --- TRR 257 ``Particle
Physics Phenomenology after the Higgs Discovery''. The authors would like to thank Robert Harlander, Fabian Lange
and  Jonas Kohnen for communication and for providing a cross check for the renormalisation constants of
the $\Delta B=0$ theory.


\appendix

\section{Basis elements for spinor structures}
\label{app::basis}

Here we list all the basis elements of spinor structures that are used to reduce tensor integrals to scalar ones. This basis is derived from the one give in Ref.~\cite{Reeck:2024iwk} and adapted to the $\db 0$ basis. We refer to~\cite{Reeck:2024iwk} for all the technical details. Explicitly, we have\footnote{A spinor structure of the form $(\overline{b}\Gamma q)(\overline{q}\Gamma b)$ is understood.}

\newcounter{mycounter}
\setcounter{mycounter}{1}
\begin{align*}
  &B_{\arabic{mycounter}}=\mathds{1}\otimes \mathds{1}\stepcounter{mycounter}\\
  &B_{\arabic{mycounter}}=\slashed{\epsilon}_q\otimes \mathds{1}\stepcounter{mycounter}\\
  &B_{\arabic{mycounter}}=\mathds{1}\otimes \slashed{\epsilon}_q\stepcounter{mycounter}\\
  &B_{\arabic{mycounter}}=\slashed{\epsilon}_q\otimes \slashed{\epsilon}_q\stepcounter{mycounter}\\
  &B_{\arabic{mycounter}}=\gamma^{\mu_{\fpeval{floor((\value{mycounter} - 1)/4)}}}\otimes \gamma_{\mu_{\fpeval{floor((\value{mycounter} - 1)/4)}}}\stepcounter{mycounter}\\
  &B_{\arabic{mycounter}}=\slashed{\epsilon}_q\gamma^{\mu_{\fpeval{floor((\value{mycounter} - 1)/4)}}}\otimes \gamma_{\mu_{\fpeval{floor((\value{mycounter} - 1)/4)}}}\stepcounter{mycounter}\\
  &B_{\arabic{mycounter}}=\gamma^{\mu_{\fpeval{floor((\value{mycounter} - 1)/4)}}}\otimes \gamma_{\mu_{\fpeval{floor((\value{mycounter} - 1)/4)}}}\slashed{\epsilon}_q\stepcounter{mycounter}\\
  &B_{\arabic{mycounter}}=\slashed{\epsilon}_q\gamma^{\mu_{\fpeval{floor((\value{mycounter} - 1)/4)}}}\otimes\gamma_{\mu_{\fpeval{floor((\value{mycounter} - 1)/4)}}} \slashed{\epsilon}_q\stepcounter{mycounter}\\
  &B_{\arabic{mycounter}}=\gamma^{\mu_1}\gamma^{\mu_{\fpeval{floor((\value{mycounter} - 1)/4)}}}\otimes \gamma_{\mu_{\fpeval{floor((\value{mycounter} - 1)/4)}}}\gamma_{\mu_1}\stepcounter{mycounter}\\
  &B_{\arabic{mycounter}}=\slashed{\epsilon}_q\gamma^{\mu_1}\gamma^{\mu_{\fpeval{floor((\value{mycounter} - 1)/4)}}}\otimes \gamma_{\mu_{\fpeval{floor((\value{mycounter} - 1)/4)}}}\gamma_{\mu_1}\stepcounter{mycounter}\\
  &B_{\arabic{mycounter}}=\gamma^{\mu_1}\gamma^{\mu_{\fpeval{floor((\value{mycounter} - 1)/4)}}}\otimes \gamma_{\mu_{\fpeval{floor((\value{mycounter} - 1)/4)}}}\gamma_{\mu_1}\slashed{\epsilon}_q\stepcounter{mycounter}\\
  &B_{\arabic{mycounter}}=\slashed{\epsilon}_q\gamma^{\mu_1}\gamma^{\mu_{\fpeval{floor((\value{mycounter} - 1)/4)}}}\otimes\gamma_{\mu_{\fpeval{floor((\value{mycounter} - 1)/4)}}}\gamma_{\mu_1} \slashed{\epsilon}_q\stepcounter{mycounter}\\
  &B_{\arabic{mycounter}}=\gamma^{\mu_1}\cdots\gamma^{\mu_{\fpeval{floor((\value{mycounter} - 1)/4)}}}\otimes \gamma_{\mu_{\fpeval{floor((\value{mycounter} - 1)/4)}}}\cdots\gamma_{\mu_1}\stepcounter{mycounter}\\
  &B_{\arabic{mycounter}}=\slashed{\epsilon}_q\gamma^{\mu_1}\cdots\gamma^{\mu_{\fpeval{floor((\value{mycounter} - 1)/4)}}}\otimes \gamma_{\mu_{\fpeval{floor((\value{mycounter} - 1)/4)}}}\cdots\gamma_{\mu_1}\stepcounter{mycounter}\\
  &B_{\arabic{mycounter}}=\gamma^{\mu_1}\cdots\gamma^{\mu_{\fpeval{floor((\value{mycounter} - 1)/4)}}}\otimes \gamma_{\mu_{\fpeval{floor((\value{mycounter} - 1)/4)}}}\cdots\gamma_{\mu_1}\slashed{\epsilon}_q\stepcounter{mycounter}\\
  &B_{\arabic{mycounter}}=\slashed{\epsilon}_q\gamma^{\mu_1}\cdots\gamma^{\mu_{\fpeval{floor((\value{mycounter} - 1)/4)}}}\otimes\gamma_{\mu_{\fpeval{floor((\value{mycounter} - 1)/4)}}}\cdots\gamma_{\mu_1} \slashed{\epsilon}_q\stepcounter{mycounter}\\
  &B_{\arabic{mycounter}}=\gamma^{\mu_1}\cdots\gamma^{\mu_{\fpeval{floor((\value{mycounter} - 1)/4)}}}\otimes \gamma_{\mu_{\fpeval{floor((\value{mycounter} - 1)/4)}}}\cdots\gamma_{\mu_1}\stepcounter{mycounter}\\
  &B_{\arabic{mycounter}}=\slashed{\epsilon}_q\gamma^{\mu_1}\cdots\gamma^{\mu_{\fpeval{floor((\value{mycounter} - 1)/4)}}}\otimes \gamma_{\mu_{\fpeval{floor((\value{mycounter} - 1)/4)}}}\cdots\gamma_{\mu_1}\stepcounter{mycounter}\\
  &B_{\arabic{mycounter}}=\gamma^{\mu_1}\cdots\gamma^{\mu_{\fpeval{floor((\value{mycounter} - 1)/4)}}}\otimes \gamma_{\mu_{\fpeval{floor((\value{mycounter} - 1)/4)}}}\cdots\gamma_{\mu_1}\slashed{\epsilon}_q\stepcounter{mycounter}\\
  &B_{\arabic{mycounter}}=\slashed{\epsilon}_q\gamma^{\mu_1}\cdots\gamma^{\mu_{\fpeval{floor((\value{mycounter} - 1)/4)}}}\otimes\gamma_{\mu_{\fpeval{floor((\value{mycounter} - 1)/4)}}}\cdots\gamma_{\mu_1} \slashed{\epsilon}_q\stepcounter{mycounter}\\
  &B_{\arabic{mycounter}}=\gamma^{\mu_1}\cdots\gamma^{\mu_{\fpeval{floor((\value{mycounter} - 1)/4)}}}\otimes \gamma_{\mu_{\fpeval{floor((\value{mycounter} - 1)/4)}}}\cdots\gamma_{\mu_1}\stepcounter{mycounter}\\
  &B_{\arabic{mycounter}}=\slashed{\epsilon}_q\gamma^{\mu_1}\cdots\gamma^{\mu_{\fpeval{floor((\value{mycounter} - 1)/4)}}}\otimes \gamma_{\mu_{\fpeval{floor((\value{mycounter} - 1)/4)}}}\cdots\gamma_{\mu_1}\stepcounter{mycounter}\\
  &B_{\arabic{mycounter}}=\gamma^{\mu_1}\cdots\gamma^{\mu_{\fpeval{floor((\value{mycounter} - 1)/4)}}}\otimes \gamma_{\mu_{\fpeval{floor((\value{mycounter} - 1)/4)}}}\cdots\gamma_{\mu_1}\slashed{\epsilon}_q\stepcounter{mycounter}\\
  &B_{\arabic{mycounter}}=\slashed{\epsilon}_q\gamma^{\mu_1}\cdots\gamma^{\mu_{\fpeval{floor((\value{mycounter} - 1)/4)}}}\otimes\gamma_{\mu_{\fpeval{floor((\value{mycounter} - 1)/4)}}}\cdots\gamma_{\mu_1} \slashed{\epsilon}_q\stepcounter{mycounter}\\
  &B_{\arabic{mycounter}}=\gamma^{\mu_1}\cdots\gamma^{\mu_{\fpeval{floor((\value{mycounter} - 1)/4)}}}\otimes \gamma_{\mu_{\fpeval{floor((\value{mycounter} - 1)/4)}}}\cdots\gamma_{\mu_1}\stepcounter{mycounter}\\
  &B_{\arabic{mycounter}}=\slashed{\epsilon}_q\gamma^{\mu_1}\cdots\gamma^{\mu_{\fpeval{floor((\value{mycounter} - 1)/4)}}}\otimes \gamma_{\mu_{\fpeval{floor((\value{mycounter} - 1)/4)}}}\cdots\gamma_{\mu_1}\stepcounter{mycounter}\\
  &B_{\arabic{mycounter}}=\gamma^{\mu_1}\cdots\gamma^{\mu_{\fpeval{floor((\value{mycounter} - 1)/4)}}}\otimes \gamma_{\mu_{\fpeval{floor((\value{mycounter} - 1)/4)}}}\cdots\gamma_{\mu_1}\slashed{\epsilon}_q\stepcounter{mycounter}\\
  &B_{\arabic{mycounter}}=\slashed{\epsilon}_q\gamma^{\mu_1}\cdots\gamma^{\mu_{\fpeval{floor((\value{mycounter} - 1)/4)}}}\otimes\gamma_{\mu_{\fpeval{floor((\value{mycounter} - 1)/4)}}}\cdots\gamma_{\mu_1} \slashed{\epsilon}_q\stepcounter{mycounter}\\
%
  &B_{\arabic{mycounter}}=\gamma^{\mu_1}\cdots\gamma^{\mu_{\fpeval{floor((\value{mycounter} - 1)/4)}}}\otimes \gamma_{\mu_{\fpeval{floor((\value{mycounter} - 1)/4)}}}\cdots\gamma_{\mu_1}\stepcounter{mycounter}\\
  &B_{\arabic{mycounter}}=\slashed{\epsilon}_q\gamma^{\mu_1}\cdots\gamma^{\mu_{\fpeval{floor((\value{mycounter} - 1)/4)}}}\otimes \gamma_{\mu_{\fpeval{floor((\value{mycounter} - 1)/4)}}}\cdots\gamma_{\mu_1}\stepcounter{mycounter}\\
  &B_{\arabic{mycounter}}=\gamma^{\mu_1}\cdots\gamma^{\mu_{\fpeval{floor((\value{mycounter} - 1)/4)}}}\otimes \gamma_{\mu_{\fpeval{floor((\value{mycounter} - 1)/4)}}}\cdots\gamma_{\mu_1}\slashed{\epsilon}_q\stepcounter{mycounter}\\
  &B_{\arabic{mycounter}}=\slashed{\epsilon}_q\gamma^{\mu_1}\cdots\gamma^{\mu_{\fpeval{floor((\value{mycounter} - 1)/4)}}}\otimes\gamma_{\mu_{\fpeval{floor((\value{mycounter} - 1)/4)}}}\cdots\gamma_{\mu_1} \slashed{\epsilon}_q\stepcounter{mycounter}\\
  &B_{\arabic{mycounter}}=\gamma^{\mu_1}\cdots\gamma^{\mu_{\fpeval{floor((\value{mycounter} - 1)/4)}}}\otimes \gamma_{\mu_{\fpeval{floor((\value{mycounter} - 1)/4)}}}\cdots\gamma_{\mu_1}\stepcounter{mycounter}\\
  &B_{\arabic{mycounter}}=\slashed{\epsilon}_q\gamma^{\mu_1}\cdots\gamma^{\mu_{\fpeval{floor((\value{mycounter} - 1)/4)}}}\otimes \gamma_{\mu_{\fpeval{floor((\value{mycounter} - 1)/4)}}}\cdots\gamma_{\mu_1}\stepcounter{mycounter}\\
  &B_{\arabic{mycounter}}=\gamma^{\mu_1}\cdots\gamma^{\mu_{\fpeval{floor((\value{mycounter} - 1)/4)}}}\otimes \gamma_{\mu_{\fpeval{floor((\value{mycounter} - 1)/4)}}}\cdots\gamma_{\mu_1}\slashed{\epsilon}_q\stepcounter{mycounter}\\
  &B_{\arabic{mycounter}}=\slashed{\epsilon}_q\gamma^{\mu_1}\cdots\gamma^{\mu_{\fpeval{floor((\value{mycounter} - 1)/4)}}}\otimes\gamma_{\mu_{\fpeval{floor((\value{mycounter} - 1)/4)}}}\cdots\gamma_{\mu_1} \slashed{\epsilon}_q\stepcounter{mycounter}\\
  &B_{\arabic{mycounter}}=\gamma^{\mu_1}\cdots\gamma^{\mu_{\fpeval{floor((\value{mycounter} - 1)/4)}}}\otimes \gamma_{\mu_{\fpeval{floor((\value{mycounter} - 1)/4)}}}\cdots\gamma_{\mu_1}\stepcounter{mycounter}\\
  &B_{\arabic{mycounter}}=\slashed{\epsilon}_q\gamma^{\mu_1}\cdots\gamma^{\mu_{\fpeval{floor((\value{mycounter} - 1)/4)}}}\otimes \gamma_{\mu_{\fpeval{floor((\value{mycounter} - 1)/4)}}}\cdots\gamma_{\mu_1}\stepcounter{mycounter}\\
  &B_{\arabic{mycounter}}=\gamma^{\mu_1}\cdots\gamma^{\mu_{\fpeval{floor((\value{mycounter} - 1)/4)}}}\otimes \gamma_{\mu_{\fpeval{floor((\value{mycounter} - 1)/4)}}}\cdots\gamma_{\mu_1}\slashed{\epsilon}_q\stepcounter{mycounter}\\
  &B_{\arabic{mycounter}}=\slashed{\epsilon}_q\gamma^{\mu_1}\cdots\gamma^{\mu_{\fpeval{floor((\value{mycounter} - 1)/4)}}}\otimes\gamma_{\mu_{\fpeval{floor((\value{mycounter} - 1)/4)}}}\cdots\gamma_{\mu_1} \slashed{\epsilon}_q\stepcounter{mycounter}\\
  &B_{\arabic{mycounter}}=\gamma^{\mu_1}\cdots\gamma^{\mu_{\fpeval{floor((\value{mycounter} - 1)/4)}}}\otimes \gamma_{\mu_{\fpeval{floor((\value{mycounter} - 1)/4)}}}\cdots\gamma_{\mu_1}\stepcounter{mycounter}\\
  &B_{\arabic{mycounter}}=\slashed{\epsilon}_q\gamma^{\mu_1}\cdots\gamma^{\mu_{\fpeval{floor((\value{mycounter} - 1)/4)}}}\otimes \gamma_{\mu_{\fpeval{floor((\value{mycounter} - 1)/4)}}}\cdots\gamma_{\mu_1}\stepcounter{mycounter}\\
  &B_{\arabic{mycounter}}=\gamma^{\mu_1}\cdots\gamma^{\mu_{\fpeval{floor((\value{mycounter} - 1)/4)}}}\otimes \gamma_{\mu_{\fpeval{floor((\value{mycounter} - 1)/4)}}}\cdots\gamma_{\mu_1}\slashed{\epsilon}_q\stepcounter{mycounter}\\
  &B_{\arabic{mycounter}}=\slashed{\epsilon}_q\gamma^{\mu_1}\cdots\gamma^{\mu_{\fpeval{floor((\value{mycounter} - 1)/4)}}}\otimes\gamma_{\mu_{\fpeval{floor((\value{mycounter} - 1)/4)}}}\cdots\gamma_{\mu_1} \slashed{\epsilon}_q\stepcounter{mycounter}\\
\end{align*}
where $\slashed{\epsilon}_q=\slashed{q} / \sqrt{q^2}$
, with $q$ being the external momentum and $q^2=m_b^2$.


\bibliographystyle{jhep}

\bibliography{paper_life.bib,extra.bib}


\end{document}